\documentclass[a4paper,11pt]{article}
\pdfoutput=1

\usepackage[DIV12]{typearea}
\usepackage{bbm}
\usepackage{amsmath}
\usepackage{amsfonts}
\usepackage{graphicx}
\usepackage{cancel}
\usepackage{booktabs}
\usepackage{multirow}
\usepackage{caption}
\usepackage{subcaption}
\usepackage{nicefrac}
\usepackage{longtable}
\usepackage{cite}

\newcommand{\rep}[1]{\ensuremath\boldsymbol{#1}}
\newcommand{\crep}[1]{\ensuremath\bar{\boldsymbol{#1}}}
\newcommand{\Z}[1]{\ensuremath{\mathbbm{Z}_{#1}}} 
\newcommand{\SO}[1]{\ensuremath{\mathrm{SO}(#1)}}
\newcommand{\SU}[1]{\ensuremath{\mathrm{SU}(#1)}}
\newcommand{\U}[1]{\ensuremath{\mathrm{U}(#1)}}
\newcommand{\E}[1]{\ensuremath{\mathrm{E}_{#1}}}

\newcommand{\nphantom}[1]{\sbox0{#1}\hspace{-\the\wd0}}

\usepackage{xcolor}

\definecolor{darkgreen}{rgb}{0.0, 0.6, 0.2}

\advance \headheight by 3.0truept       
\usepackage[pdftex]{hyperref}
\hypersetup{bookmarksnumbered,colorlinks,
    linkcolor={black},
    citecolor={black},
    urlcolor={black}}

\hypersetup{
    pdftitle = {Contrast data mining for the MSSM from strings},
    pdfauthor = {Parr, Vaudrevange}
}
 
\begin{document}

\begin{titlepage}
\vspace*{-1cm}
\begin{flushright}
TUM-HEP 1233/19\\
\end{flushright}

\vspace*{0.5cm}

\begin{center}
{\Huge\textbf{
Contrast data mining for the MSSM from strings \\
}
}

\vspace{0.7cm}

\textbf{Erik Parr and Patrick K.S. Vaudrevange
}
\\[8mm]
\textit{\small
~Physik Department T75, Technische Universit\"at M\"unchen, \\
James--Franck--Stra\ss e 1, 85748 Garching, Germany}
\end{center}

\vspace{0.4cm}

\begin{abstract}
We apply techniques from data mining to the heterotic orbifold landscape in order to identify new 
MSSM-like string models. To do so, so-called contrast patterns are uncovered that help to 
distinguish between areas in the landscape that contain MSSM-like models and the rest of the 
landscape. First, we develop these patterns in the well-known $\mathbbm{Z}_6$-II orbifold 
geometry and then we generalize them to all other $\mathbbm{Z}_N$ orbifold geometries. Our contrast 
patterns have a clear physical interpretation and are easy to check for a given string model. 
Hence, they can be used to scale down the potentially interesting area in the landscape, which 
significantly enhances the search for MSSM-like models. Thus, by deploying the knowledge gain from 
contrast mining into a new search algorithm we create many novel MSSM-like models, especially in 
corners of the landscape that were hardly accessible by the conventional search algorithm, for 
example, MSSM-like $\mathbbm{Z}_6$-II models with $\Delta(54)$ flavor symmetry.
\end{abstract}

\end{titlepage}

\section{Introduction}  
String theory is a promising candidate for a UV-complete theory of quantum gravity. However, to 
proof its usefulness it has to incorporate the standard model (SM) or its supersymmetric extension 
(the MSSM) as a low-energy limit. Thus, one of the main tasks of string phenomenology is to find a 
string model that is consistent with all experimental and observational data -- or, at least, that 
comes close to the MSSM, i.e.\ a model that is MSSM-like. This task is very 
difficult, mainly due to two reasons: (i) String theory predicts extra spatial dimensions. Hence, 
the connection between string theory and the MSSM must be related to the way how the extra 
dimensions are compactified. However, the number of different compactifications is 
huge~\cite{Lerche:1986cx,Douglas:2003um}, giving rise to the so-called string landscape of 
effective 4D theories originating from string theory. (ii) String theory is very predictive because 
after specifying the compactification the effective 4D theory, including all symmetries, particles 
and couplings, is completely fixed.

In the case of the ten-dimensional $\E{8}\times\E{8}$ heterotic string we have to compactify six 
dimensions. To do so, we choose six-dimensional toroidal orbifolds~\cite{Dixon:1985jw,Dixon:1986jc} 
as they allow for a world-sheet formulation of string theory instead of a supergravity approximation, 
see e.g.~\cite{Faraggi:1992fa,Dijkstra:2004cc,Braun:2005ux,Gmeiner:2005vz,Dienes:2006ut,Blumenhagen:2006ci,Anderson:2011ns,Anderson:2012yf,Cvetic:2015txa,Cvetic:2018ryq,Cvetic:2019gnh} 
for other schemes. Then, the conventional approach to search for MSSM-like string models from 
heterotic orbifolds is given by a random scan in the 
landscape~\cite{Nibbelink:2013lua,Nilles:2014owa,Olguin-Trejo:2018wpw} or in fertile islands that 
can be identified by certain patterns, like local GUTs~\cite{Lebedev:2006kn,Lebedev:2008un,Pena:2012ki}. 
In this paper, we show that the approach of defining fertile islands can be generalized by applying 
machine learning techniques to the string landscape~\cite{He:2017aed,Krefl:2017yox,Ruehle:2017mzq,Carifio:2017bov}, 
see also~\cite{Wang:2018rkk,Bull:2018uow,Klaewer:2018sfl,Halverson:2019tkf,Cole:2018emh,Cole:2019enn,Bull:2019cij,ashmore2019machine}. 
A first hint towards such a generalization was observed in ref.~\cite{Mutter:2018sra}: a neural 
network was able to identify additional patterns and to cluster the models of the heterotic orbifold 
landscape according to them. Surprisingly, MSSM-like models turned out to be localized 
close to each other on fertile islands, even though the neural network did not know whether a given 
model is MSSM-like or not, denoted by \cancel{MSSM}-like.

In this paper we propose and demonstrate a new search strategy for MSSM-like models based on 
additional patterns that is superior by orders of magnitude. This search is developed from the 
knowledge gained by analyzing the heterotic orbifold landscape with tools from data mining. Data 
mining has been developed for the purpose to prepare, visualize and explore huge databases. Hence, 
the suitability of data mining to the string landscape is obvious. In particular, we apply a 
special setup called contrast data mining to the heterotic orbifold landscape. The basic idea of 
contrast data mining is to identify so-called contrast patterns that allow us to focus our search 
on those areas in the landscape where the MSSM-like models reside. Our contrast patterns have a 
clear physical interpretation: they are given by the number of unbroken roots in the hidden $\E{8}$ 
factor~\cite{Dienes:2006ut,Lebedev:2006tr} and the numbers of various bulk matter fields. 

This paper is structured as follows: In section~\ref{sec:setup_random_search} we review the 
conventional search algorithm for MSSM-like models in the heterotic orbifold landscape. Then, we 
improve this algorithm using traditional methods: first, we take the Weyl symmetry into account 
and, secondly, in section~\ref{sec:pheno} we include phenomenological constraints. These 
improvements already reduce the number of models that have to be scanned by the search algorithm in 
the test case of the $\Z{6}$-II orbifold geometry. Afterwards, in section~\ref{sec:contrast_pattern} 
we apply data mining techniques to our $\Z{6}$-II dataset. Doing so, we can identify contrast 
patterns that help to distinguish between areas in the landscape that contain MSSM-like models and 
others. We implement these contrast patterns into our search algorithm and show that we can easily 
construct many new MSSM-like $\Z{6}$-II models -- a fact that might be surprising given the 
extensive searches done especially in this orbifold geometry. Remarkably, we can identify two 
MSSM-like $\Z{6}$-II models with $\Delta(54)$ flavor symmetry (related to a vanishing Wilson line 
of order three), see section~\ref{sec:delta54}. Thereafter, our contrast patterns are successfully 
extended to all $\Z{N}$ orbifold geometries in section~\ref{sec:geometry_dependent}, where 
table~\ref{tab:inequiv_ZN} summarizes our results, see also~\cite{Parr:2019anc}. Finally, in 
section~\ref{sec:conclusion} we conclude.

\begin{table}[ht!]
\center
\begin{tabular}{ccc}
vector  & order $N_k$ & additional constraint \\ 
\midrule
$V_{1}$ &  6    & \\
$V_{2}$ &  1    & not present, i.e. $V_2 = (0^{16})$\\
\midrule[0.15mm]
$W_{1}$ &  1    & $W_{1}=(0^{16})$ \\
$W_{2}$ &  1    & $W_{2}=(0^{16})$ \\
\midrule[0.15mm]
$W_{3}$ &  3    & \multirow{2}{*}{$W_{3} = W_{4}$}\\
$W_{4}$ &  3    & \\
\midrule[0.15mm]
$W_{5}$ &  2    & \\
$W_{6}$ &  2    & \\
\end{tabular}
\caption{\label{tab:geometry_constraints}
Table of geometrical constraints for shift vectors and Wilson lines in the case of the $\Z{6}$-II 
orbifold geometry.}
\end{table}

\section{Searching the heterotic orbifold landscape}
\label{sec:setup_random_search}

An orbifold compactification~\cite{Dixon:1985jw,Dixon:1986jc} is specified by a six-dimensional 
orbifold geometry $\mathbbm{O}$ and a gauge embedding that acts on the sixteen gauge degrees of 
freedom of the heterotic string. While there exist only 138 orbifold geometries with Abelian point 
group (i.e.\ $\Z{N_1}$ or $\Z{N_1}\times\Z{N_2}$) and $\mathcal{N}=1$ supersymmetry~\cite{Fischer:2012qj}, 
the true size of the heterotic orbifold landscape unfolds if we take the number of inequivalent 
gauge embeddings into account. For a general $\Z{N_1}\times\Z{N_2}$ orbifold, a gauge embedding is 
given by two shift vectors, $V_{1}$ and $V_{2}$, and six Wilson lines, $W_{1}$ to $W_{6}$, corresponding 
to the six compactified directions. In this paper we concentrate on the $\E{8} \times \E{8}$ 
heterotic string\footnote{The methods described in this paper are easy to generalize to the \SO{32} 
heterotic string.}, which implies that each vector can be split into two eight-dimensional vectors. 
For example, the sixteen-dimensional shift vector $V_{1}$ consists of the eight-dimensional vectors 
$V^{(1)}_{1}$ and $V^{(2)}_{1}$, which act on the first and second $\E{8}$ factor, respectively. 
Altogether a gauge embedding is determined by a gauge embedding matrix
\begin{equation}
M~=~
\begin{pmatrix}
V^{(1)}_{1} & V^{(2)}_{1} \\
V^{(1)}_{2} & V^{(2)}_{2} \\
W^{(1)}_{1} & W^{(2)}_{1} \\
W^{(1)}_{2} & W^{(2)}_{2} \\
W^{(1)}_{3} & W^{(2)}_{3} \\
W^{(1)}_{4} & W^{(2)}_{4} \\
W^{(1)}_{5} & W^{(2)}_{5} \\
W^{(1)}_{6} & W^{(2)}_{6} \\
\end{pmatrix}\;,
\end{equation}
where we denote the vector in the $k$-th line by $M_{k}$ for $k=1,\ldots,8$ and split it into two 
parts $M_{k}^{(\alpha)}$ for $\alpha = 1,2$ corresponding to the two $\E{8}$ factors, for example, 
$M_{3} = W_1 = (W^{(1)}_{1},W^{(2)}_{1})$ for the first Wilson line $W_1$. Depending on the 
orbifold geometry, shift vectors and Wilson lines are subject to geometrical constraints that, for 
example, fix the order of the shift vector $V_1$ to $N$ for a $\Z{N}$ orbifold geometry, see e.g. 
ref.~\cite{RamosSanchez:2008tn}. To be more specific and as we will mainly work with the so-called 
$\Z{6}$-II $(1, 1)$ orbifold geometry (using the nomenclature from ref.~\cite{Fischer:2012qj}, 
abbreviated as $\Z{6}$-II in the following), we summarize the geometrical constraints on the shift 
vectors and Wilson lines for the $\Z{6}$-II orbifold geometry in table~\ref{tab:geometry_constraints}.

In general, there are two ways to expand a sixteen-dimensional shift vector or Wilson line 
naturally: either in terms of the simple roots $\alpha_{I}$ of $\E{8}\times\E{8}$ or in terms of 
the dual simple roots $\alpha^{*}_{I}$, $I=1,\ldots,16$, where 
$\alpha^{*}_{I} \cdot \alpha_{J} = \delta_{IJ}$. Both choices give a basis of the (self-dual) root 
lattice $\Lambda_{\E{8}\times\E{8}}$ of $\E{8}\times\E{8}$. For later convenience (see 
section~\ref{sec:WeylSymmetry}) we decide to expand the vectors in terms of the dual basis, i.e.\ 
we parameterize the vectors $M_{k}$ in the gauge embedding matrix $M$ as
\begin{equation}\label{eq:ExpansionInDualBasis}
M_{k} ~=~ \dfrac{1}{N_k}\sum_{I=1}^{16} \ d_{k\,I}  \ \alpha^{*}_{I}\;.
\end{equation}
Here, $N_k$ defines the order of the shift vector or Wilson line and $d_{k\,I} \in \mathbb{Z}$ for 
$k=1,\ldots,8$ and $I=1,\ldots,16$ are integers. Consequently, a gauge embedding matrix $M$ 
corresponds to a point in $d \in \Z{}^{128}$ since $d_{k\,I}$ has $8 \times 16 = 128$ components. 
Note that this construction eq.~\eqref{eq:ExpansionInDualBasis} inherently ensures the correct 
order of the respective vector. In detail, for all vectors $k=1,\ldots,8$ we have
\begin{equation}
N_k\, M_{k} ~\in~ \Lambda_{\E{8}\times\E{8}}\;.
\end{equation}

\subsection{Conditions from modular invariance}
\label{sec:modular_invariance}

\begin{figure}[t!]
\includegraphics[width=10cm]{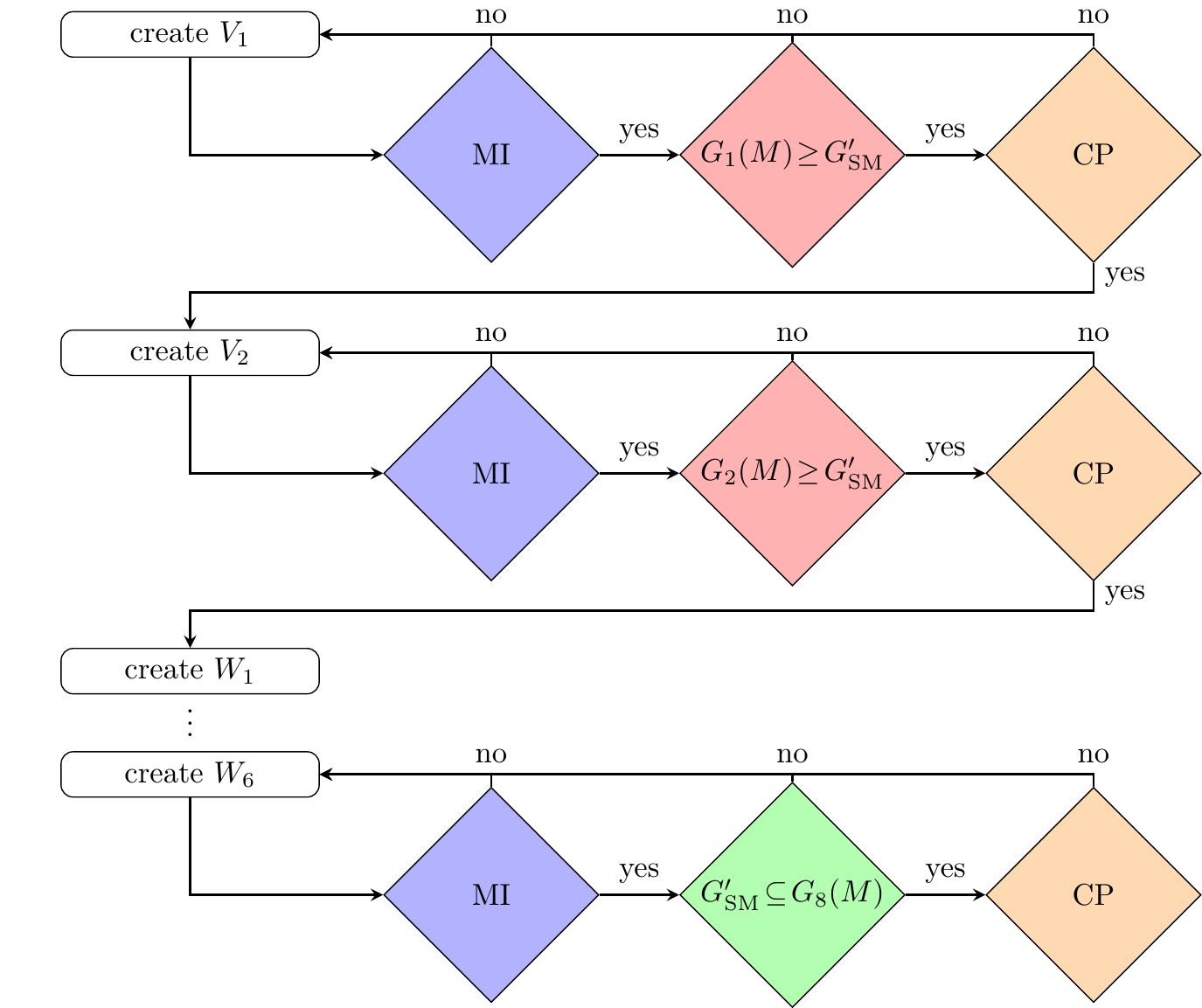}
\centering
\caption{Flowchart of the construction of shift vectors and Wilson lines, starting with 
the shift vector $V_1$ at step $n=1$ and ending with the Wilson line $W_6$ at step $n=8$. At each 
step $n=1,\ldots,8$, the vector $M_n$ is chosen randomly and the corresponding modular invariance 
(MI) conditions are tested. If the vector passes this test, two additional conditions are applied:\\
(i) As discussed in section~\ref{sec:constraint_GUT} the gauge group $G_n(M)$ is computed using the 
already chosen vectors $M_k$ for $k=1,\ldots,n$. If $G_n(M)$ satisfies a necessary condition to 
host the non-Abelian gauge group factors $G_{\mathrm{SM}}' = \SU{3}\times\SU{2}$ of the SM in the 
first $\E{8}$ factor, i.e. $G_n(M) \geq G_{\mathrm{SM}}'$ in terms of their root systems, the model 
is passed on to the next condition. \\
(ii) As introduced in section~\ref{sec:contrast_pattern}, the contrast patterns (CP) are imposed.\\
Finally, after the last Wilson line $M_8 = W_6$ has been chosen successfully, the four-dimensional 
gauge group $G_{4D}(M) = G_8(M)$ must contain $G_{\mathrm{SM}}'$ in the first $\E{8}$ as described 
in section~\ref{sec:constraint_G_sm}.}
\label{fig:successive_create}
\end{figure}

In order to obtain consistent string compactifications we have to impose conditions from modular 
invariance of the one-loop string partition function on the gauge embedding matrix $M$. These 
conditions read~\cite{Ploger:2007iq}
\begin{subequations}\label{eq:modularinvariance}
\begin{eqnarray}
N_1\,\left(V_1^2 - v_1^2 \right)                                 & = & 0 \mod 2\;, \\
N_2\,\left(V_2^2 - v_2^2 \right)                                 & = & 0 \mod 2\;, \\
\mathrm{gcd}(N_1, N_2)\,\left(V_1\cdot V_2 - v_1\cdot v_2 \right)& = & 0 \mod 2\;, \\
\mathrm{gcd}(N_{i+2}, N_1)\,\left(W_i\cdot V_1\right)            & = & 0 \mod 2\;, \\
\mathrm{gcd}(N_{i+2}, N_2)\,\left(W_i\cdot V_2\right)            & = & 0 \mod 2\;, \\
N_{i+2}\,\left(W_i^2\right)                                      & = & 0 \mod 2\;, \\
\mathrm{gcd}(N_{i+2},N_{j+2})\,\left(W_i\cdot W_j\right)         & = & 0 \mod 2 \quad (i \neq j)\;,
\end{eqnarray}
\end{subequations}
for $i,j=1,\ldots,6$ and where $v_{1}$ and $v_{2}$ denote the so-called twist vectors. They encode 
the geometrical rotation angles of a general $\Z{N_1} \times \Z{N_2}$ orbifold geometry, while 
for \Z{N_1} orbifolds the twist $v_{2}=(0^4)$ is not present. In addition, the gcd in 
eq.~\eqref{eq:modularinvariance} denotes the greatest common divisor. These conditions are very 
restrictive and already forbid a huge fraction of points in the space $\Z{}^{128}$ corresponding to 
eq.~\eqref{eq:ExpansionInDualBasis}. The only reasonable way to create a consistent gauge embedding 
matrix $M$ is by successively creating shift vectors and Wilson lines step-by-step and checking 
each time if the relevant conditions from eqs.~\eqref{eq:modularinvariance} are fulfilled for those 
combinations of shift vectors and Wilson lines that have been chosen so far, see 
figure~\ref{fig:successive_create}.

In this paper we work out an extension of this logic, i.e.\ we create a successive search that only 
considers those areas in the heterotic orbifold landscape that can fulfill certain properties: 
first, in section~\ref{sec:pheno} we will introduce phenomenological properties of the MSSM and 
then, in the main part of the paper in section~\ref{sec:contrast_pattern}, we define so-called 
contrast patterns that also can be checked at each step during the construction of a gauge 
embedding matrix $M$. By doing so, we will neglect those areas in the heterotic orbifold landscape 
that have no chance or an extremely low probability to host a valid MSSM-like orbifold model.

\newpage 
\subsection{The Orbifolder}
\label{sec:orbifolder}

Given an orbifold geometry $\mathbbm{O}$ and a consistent gauge embedding matrix $M$, it is in 
principle possible to compute the complete 4D orbifold model at low energies denoted by 
model$(M)$.~\footnote{If the orbifold geometry $\mathbbm{O}$ is clear from the context, we will 
also name $M$ as orbifold model.} In practice, some computations are too complicated, e.g.\ the 
strengths of Yukawa couplings. However, for a given $M$ one can use the 
\texttt{orbifolder}~\cite{Nilles:2011aj}~\footnote{All 138 orbifold geometries with Abelian point 
groups and $\mathcal{N}=1$ supersymmetry~\cite{Fischer:2012qj} are encoded in so-called 
geometry-files that can be read by the \texttt{orbifolder}. These geometry-files can be found as 
ancillary files to ref.~\cite{Ramos-Sanchez:2018edc}.} in order to get the massless string 
spectrum, denoted by spectrum$(M)$, with all gauge charges. Moreover, the \texttt{orbifolder} can 
identify MSSM-like models, i.e.\ models with $\SU{3}_\mathrm{C}\times\SU{2}_\mathrm{L}\times\U{1}_Y$ 
gauge symmetry and the exact chiral spectrum of the MSSM plus at least one Higgs-pair and exotics 
that have to be vector-like with respect to the SM. Also discrete symmetries and a list of all 
allowed couplings up to a certain order in fields can be analyzed. In addition, the 
\texttt{orbifolder} can be used to identify inequivalent orbifold models by taking two orbifold 
models, model$(M)$ and model$(M')$, to be equivalent if their massless spectra coincide (on the 
level of the non-Abelian representations and, in case of MSSM-like models, the $\U{1}_Y$ 
hypercharge), i.e.
\begin{equation}
\mathrm{spectrum}(M) ~=~ \mathrm{spectrum}(M') \quad\Rightarrow\quad \mathrm{model}(M) ~\sim~ \mathrm{model}(M')\;.
\end{equation}

\subsection{Searching in the Weyl chambers}
\label{sec:WeylSymmetry}

A Weyl reflection of the gauge embedding vector $M_k$ at the hyperplane orthogonal to the simple 
root $\alpha_I$ of $\E{8}\times\E{8}$ is defined as
\begin{equation}
w_{I}(M_k) ~=~ M_k - (M_k\cdot \alpha_I)\, \alpha_I\;,
\end{equation}
using $(\alpha_I)^2 = 2$ for $I=1,\ldots,16$. Then, it is easy to show that 
\begin{equation}
w_{I}(M_k) \cdot w_{I}(M_\ell) ~=~ M_k\cdot M_\ell\;.
\end{equation}
Hence, Weyl reflections leave the modular invariance conditions~\eqref{eq:modularinvariance} 
invariant. Furthermore, one can show that they are symmetries of the full string theory
\begin{align}
M' ~=~ w_I(M) \quad\Rightarrow\quad \mathrm{model}\big(M\big) ~=~ \mathrm{model}\big(M'\big)\;,
\end{align}
where $w_I$ acts simultaneously on all shift vectors and Wilson lines encoded in $M$. Hence, the 
gauge embedding matrices $M$ and $M'= w_I(M)$ are equivalent for all Weyl reflections. Now, Weyl 
reflections generate a group, the so-called Weyl group. For $\E{8}\times\E{8}$, it has 
$\approx 5 \cdot 10^{17}$ elements. Consequently, the Weyl group of $\E{8}\times\E{8}$ yields a huge 
redundancy between physically equivalent models in the heterotic orbifold landscape.

We can reduce the search space and therefore find more physically inequivalent models when we 
divide out this symmetry. For this task we propose a \emph{fundamental Weyl chamber} search. The 
proposed technique is based on the algorithm of ref.~\cite{Fuchs:1997jv} that any vector in the 
root space can be rotated to the fundamental Weyl chamber, which is defined as the subspace where 
all Dynkin labels $M_k \cdot \alpha_I$ are non-negative.

Starting from a gauge embedding matrix $M$ we can imagine to apply the algorithm of 
ref.~\cite{Fuchs:1997jv} such that the shift vector $V_{1}$ is rotated to the fundamental Weyl 
chamber, i.e.\ $V_1 \cdot \alpha_I \geq 0$ for all simple roots $I=1,\ldots,16$. Since we do not 
want to change the orbifold model by this transformation, we have to act with the same Weyl 
reflections that mapped $V_{1}$ to the fundamental Weyl chamber on the other vectors simultaneously. 
After this, we might still have some Weyl symmetries left, i.e.\ the shift vector $V_{1}$ may 
be invariant under certain Weyl reflections. These unbroken Weyl reflections are those that leave 
$V_{1}$ invariant, i.e.\ $w_{I}(V_1) = V_1$ if and only if $V_1 \cdot \alpha_I = 0$, i.e.\ 
$d_{1\,I}=0$. These residual Weyl reflections can now be used to bring the next vector, in our case 
the Wilson line $W_{1}$, to an enlarged Weyl chamber which we define in analogy to the fundamental 
Weyl chamber but using only those Weyl reflections that leave $V_{1}$ invariant. Consequently, 
after the transformation of the Wilson line $W_{1}$ to the enlarged Weyl chamber, those Dynkin 
labels $W_{1} \cdot \alpha_I$ are constrained to be non-negative that correspond to the Weyl 
reflections $w_I$ that leave the shift vector $V_{1}$ invariant. This procedure can be reapplied to 
the next vectors until no Weyl symmetry is left.  

\begin{figure*}[t!]
    \centering
    \begin{subfigure}[t]{0.45\textwidth}
        \centering
        \includegraphics[width=\textwidth]{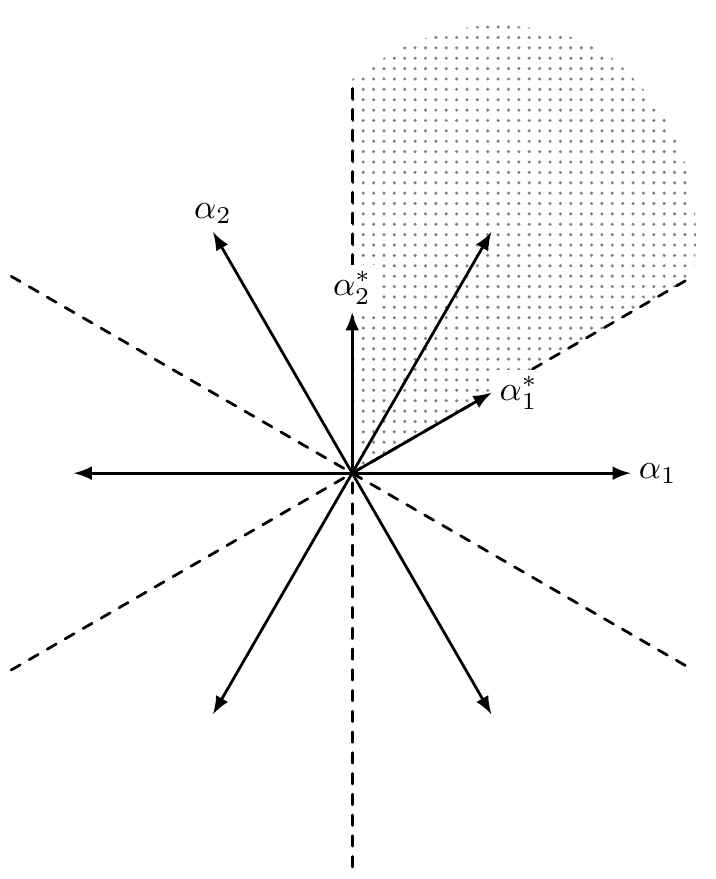}
        \caption{\label{fig:FWC:a}
        Fundamental Weyl chamber of \SU3.}
    \end{subfigure}%
    ~ 
    \begin{subfigure}[t]{0.45\textwidth}
        \centering
        \includegraphics[width=\textwidth]{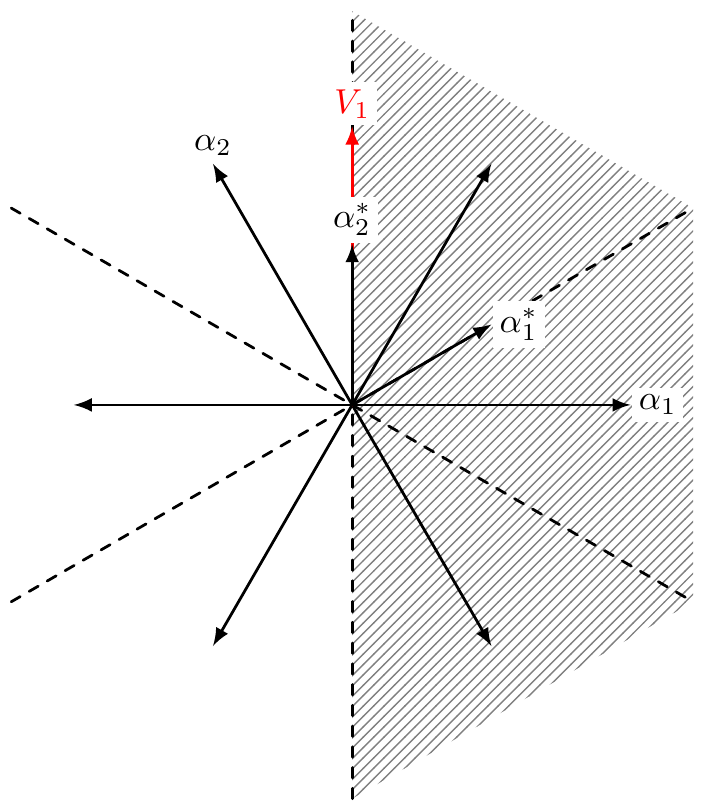}
        \caption{\label{fig:FWC:b}
        Domain for $W_{1}$ after a specific $V_{1}$ was chosen.}
    \end{subfigure}
    \caption{\label{fig:FWC}
These figures illustrate the algorithm to divide out the Weyl symmetry from the search process, 
exemplified at the root space of \SU3. In figure~\ref{fig:FWC:a} the first vector $V_1$ can be 
restricted to lie in the fundamental Weyl chamber of \SU3 (shaded area) defined by the Weyl 
reflections $w_1$ and $w_2$. Thus, $d_{1\,I} \in \mathbbm{N}_0$ for $I=1,2$. In 
figure~\ref{fig:FWC:b} we have chosen a specific vector $V_1$ along the direction of $\alpha_2^*$ 
as an example. Consequently, the vector $V_{1}$ is invariant under the Weyl reflection $w_1$, i.e.\ 
$w_1(V_1) = V_1$, which corresponds to a vanishing Dynkin label $V_1 \cdot \alpha_{1} = 0$. Hence, 
this choice for $V_{1}$ has not broken the whole Weyl symmetry and we can use the unbroken Weyl 
reflection $w_1$ to restrict the search space for $W_{1}$ to the enlarged Weyl chamber (shaded 
area) which is defined by $W_1 \cdot \alpha_1 \geq 0$. Hence, the coefficients of $W_1$ can be 
constrained as $d_{3\,1} \in \mathbbm{N}_0$ and $d_{3\,2} \in \mathbbm{Z}$. This procedure is 
continued for the next vectors and takes at each step all previously chosen vectors into account 
for computing the respective unbroken Weyl symmetry.}
\end{figure*}

The mindset above can be used to directly choose only gauge embedding matrices that solely have the 
first vector in the fundamental Weyl chamber and the following vectors are in the correspondingly 
enlarged versions of it, as illustrated in figure~\ref{fig:FWC}.
To achieve this we expand our vectors not in the basis of the simple roots $\alpha_I$ but in the 
dual basis $\alpha_I^*$, see eq.~\eqref{eq:ExpansionInDualBasis}, where we can apply the 
constraints on the Dynkin labels directly via the coefficients $d_{k\,I}$ of the gauge 
embedding matrix. For the first vector, i.e.\ the shift vector $V_{1}$, we have the full freedom of 
the Weyl group and can therefore choose this vector directly from the fundamental Weyl chamber
\begin{equation}
V_1 \cdot \alpha_I ~=~ \frac{d_{1\,I}}{N_{1}} ~\geq~ 0 \quad\Leftrightarrow\quad d_{1\,I} ~\in~ \mathbbm{N}_0\;,
\end{equation}
for $I=1,\ldots,16$. Thereafter, we have to compute the unbroken Weyl symmetry that can be 
exploited to restrict the second vector. This unbroken Weyl symmetry is generated by those Weyl 
reflections that leave $V_1$ invariant, i.e.\ $V_1$ has to be a fixed point of a Weyl reflection 
such that this Weyl reflection remains unbroken. Since we have chosen $V_1$ from the fundamental 
Weyl chamber it can only be a fixed point under a Weyl reflection if $V_1$ lies on the boundary of 
the fundamental Weyl chamber. This boundary is given by the union of the hyperplanes perpendicular 
to the simple roots, i.e.\ the hyperplanes at which the Weyl reflections $w_{I}$ 
act~\cite{Fuchs:1997jv}. Consequently, only those Weyl reflections $w_{I}$ which leave all 
previously chosen vectors $M_{k}$ invariant can still restrict the search space of the shift vector 
and Wilson lines that have to be chosen next. Therefore, at step $n$ in 
figure~\ref{fig:successive_create} we can constrain the coefficients $d_{n\,I}$ of the vector $M_n$ 
in eq.~\eqref{eq:ExpansionInDualBasis} as
\begin{subequations}
\begin{eqnarray}
d_{n\,I} & \in \mathbbm{N}_0            \quad\text{ if }& d_{k\,I}~=~0    \quad\text{for all } k=1,\dots, n-1\;,\\
d_{n\,I} & \in \mathbbm{Z}_{\phantom{0}}\quad\text{ if }& d_{k\,I}~\neq~0 \quad\text{for any } k=1,\dots, n-1\;.
\end{eqnarray}
\end{subequations}

\begin{table}[ht!]
\renewcommand{\arraystretch}{1.2}
\center
\begin{tabular}{|r|r|l|}
\hline
\multirow{2}{*}{position} & \multicolumn{1}{c|}{frequency}     & \multirow{2}{*}{type of model} \\
                          & \multicolumn{1}{c|}{of occurrence} & \\
\hline \hline
          1 & 8\,008 & gauge group $\U{1}^{16}$ (with $218$ matter fields)\\
     \vdots & \vdots & \\
         19 & 1\,915 & first non-Abelian gauge group (gauge group $\SU{2}\times\U{1}^{15}$)\\
     \vdots & \vdots & \\
     3\,700 &    114 & first gauge group $\SU{3}\times\SU{2}$\\
     \vdots & \vdots & \\
   119\,911 &     10 & first MSSM-like model with 1 generation plus vector-like exotics\\
     \vdots & \vdots & \\
1\,097\,248 &      2 & first MSSM-like model with 2 generations plus vector-like exotics\\
     \vdots & \vdots & \\
3\,560\,178 &      1 & first MSSM-like model with 3 generations plus vector-like exotics\\
\hline
\end{tabular}
\caption{$10^{7}$ random models from the $\Z{6}$-II orbifold geometry, ordered by their frequency 
of occurrence. We list only the first appearances of some special models according to the 
properties given in the last column. Note that the last position in this list would be 
$3\,690 \,513$ as this is the number of inequivalent models in this dataset.}
\label{tab:spec_MI}
\end{table}

\section{Phenomenological constraints}
\label{sec:pheno}

\begin{figure}[ht!] 
\includegraphics[width=\textwidth]{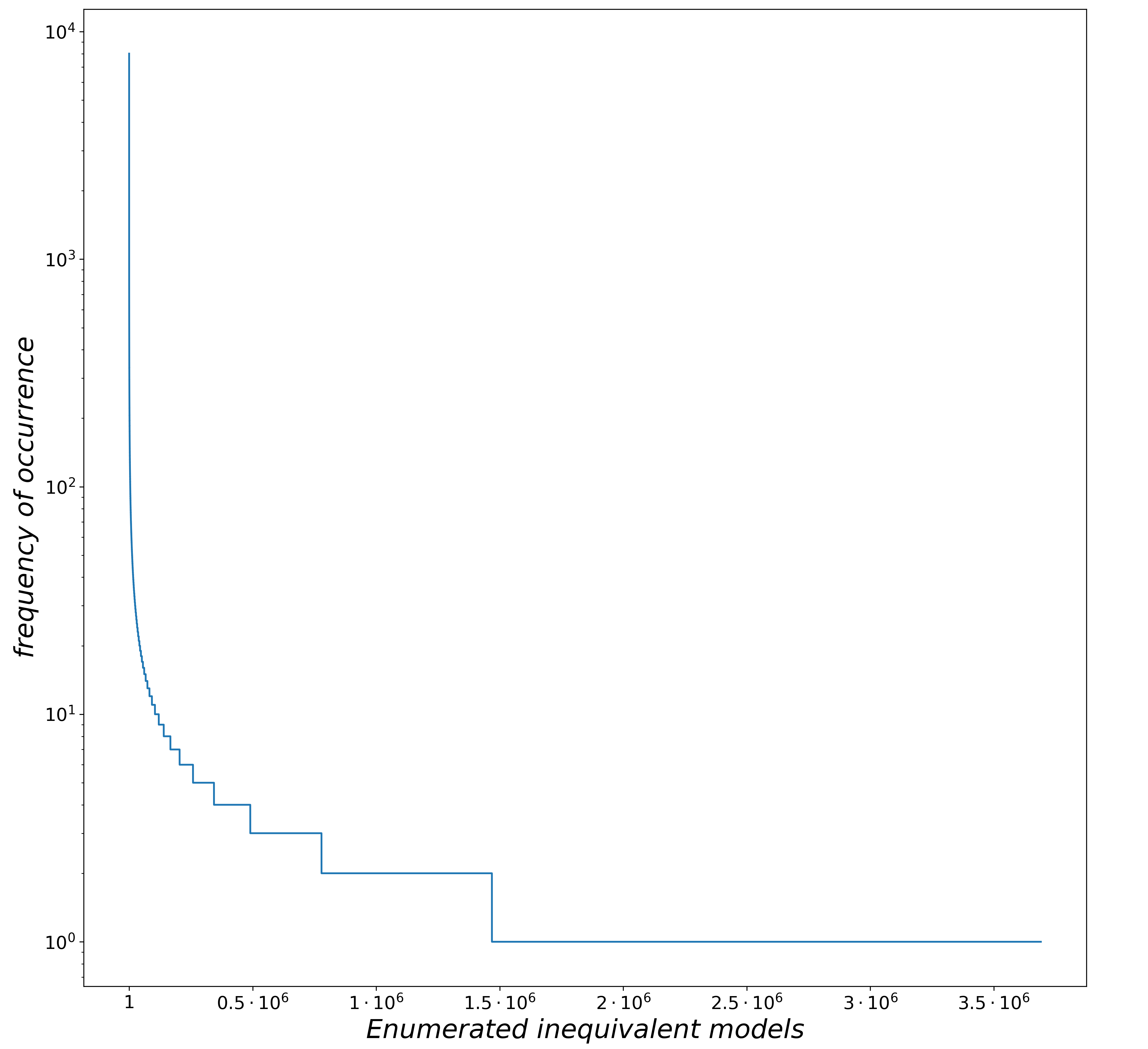}
\centering
\caption{Logarithmic plot of the frequency of occurrence of inequivalent $\Z{6}$-II orbifold 
models. On the horizontal axis, the inequivalent models are enumerated from $1$ to $3\,690 \,513$, 
while on the vertical axis we see the corresponding frequency of occurrence, i.e.\ model \# 1 has a 
frequency of $8\,008$, see also table~\ref{tab:spec_MI} for more details on some of these models.}
\label{fig:hist_spec_MI}
\end{figure}

Obviously, we have to neglect any orbifold model specified by a gauge embedding matrix $M$ that 
does not obey the stringy consistency conditions on $M$: the geometrical constraints and the 
modular invariance conditions. Similarly, we can add phenomenologically inspired constraints on $M$: 
Any orbifold model whose four-dimensional gauge symmetry $G_{4D}(M)$ does not contain the one of 
the SM does not provide a valid model to describe nature. Importantly, if we search in the 
heterotic orbifold landscape taking only the stringy consistency condition into account, these 
phenomenologically uninteresting models build by far the main part of the heterotic orbifold 
landscape. We have verified this by constructing $10^7$ random models in the $\Z{6}$-II orbifold 
geometry that satisfy all stringy consistency conditions using the \emph{fundamental Weyl chamber} 
search algorithm. These $10^7$ models give rise to approximately $3.5 \cdot 10^6$ inequivalent 
massless spectra. Then, the inequivalent spectra are sorted according to their frequency of 
occurrence inside the full list of $10^7$ random models. The result is shown in 
figure~\ref{fig:hist_spec_MI} and details on some of the inequivalent spectra are highlighted in 
table~\ref{tab:spec_MI}.
Consequently, from a phenomenological point of view the most uninteresting models turn out to have 
the highest repetition values, and the most interesting models are the rarest. As a remark, we 
cannot explain this imbalance, for example, by $\E{8}\times\E{8}$ lattice translations and Weyl 
reflections. Since the models are compared on the level of their massless spectra, it is likely 
that a lot of these models actually differ by some additional model parameters, for instance by 
their Yukawa couplings. Nevertheless, we want to avoid these uninteresting models in our search for 
MSSM-like orbifold models. Therefore, we will describe in the upcoming sections how we can constrain 
our search to those areas of the heterotic orbifold landscape that can potentially host the SM.

\subsection[The pseudo-GUT constraint Gn(M) >= GSM']{\boldmath The pseudo-GUT constraint $G_n(M) \geq G_\mathrm{SM}'$\unboldmath}
\label{sec:constraint_GUT}

We want to avoid to produce phenomenologically uninteresting models that have a gauge group smaller 
than the SM gauge group factors $G_\mathrm{SM}' = \SU{3}\times\SU{2}$. \footnote{Note that the 
existence of an anomaly-free $\U{1}_Y$ hypercharge can and will be tested only at the very end of 
the search algorithm by the \texttt{orbifolder}, after the full orbifold model has been specified.} 
Hence, we can check how far the gauge group $G_n(M)$ is already broken at each step $n = 1,\ldots, 8$ 
in the algorithm illustrated in figure~\ref{fig:successive_create}. Since additional shift vectors 
or Wilson lines can only break the gauge group further, i.e.\ $G_{n+1}(M) \subseteq G_n(M)$, the 
SM gauge group provides us with a lower bound on the breaking pattern at each step $n$. A fast way 
to check the size of the remaining gauge group $G_n(M)$ is to compute the number of unbroken roots 
$N_{\mathrm{sr}}$ after the first $n$ vectors $M_k$, for $k=1,\ldots,n$, have been chosen to 
specify a consistent gauge embedding matrix $M$,
\begin{align}\label{eq:number_surviving_roots}
& N_{\mathrm{sr}}^{(\alpha)}(M)~=~ \sum_{p \,\in\, \Phi_{\E{8}}} \left( \prod\limits_{k\,=\,1}^{n} \delta\left( p \cdot M_{k}^{(\alpha)}\right)\right)\;,
\end{align}
for $\alpha=1,2$ corresponding to both $\E{8}$ factors and $\Phi_{\E{8}}$ is defined as the root 
system of $\E{8}$ with 240 roots $p$. Furthermore, $\delta(x) = 1,0$ depending on whether $x$ is 
integer or not, respectively. In the case of the SM we have six unbroken roots for \SU3 and two 
unbroken roots for the \SU2 factor, i.e.\ $N_{\mathrm{sr}}(\mathrm{SM})=8$. This is our lower bound 
at each step $n$ in the production of our model $M$ for the first $\E{8}$ factor, i.e.
\begin{equation}\label{eq:pseudoGUTConstraint}
N_{\mathrm{sr}}^{(1)}(M) ~\geq~ 8 \quad\mathrm{at\ each\ step}\ n \;.
\end{equation}
On the other hand, the second $\E{8}$ factor is free to produce any additional hidden gauge groups. 
Note that $\SU2 \times \SU2 \times \SU2 \times \SU2$ could fulfill the 
constraint~\eqref{eq:pseudoGUTConstraint} but is already broken too far. Due to this we demand in 
addition that one gauge factor of $G_n(M)$ has a root system that allows for \SU3, i.e.\ with six 
or more unbroken roots. If a newly chosen vector $M_k$ results in a gauge group breaking below 
these lower bounds, we neglect this vector and choose the same vector again, until it fulfills this 
constraint. In the following we call this constraint the \emph{pseudo-GUT} constraint.

\subsection[The Standard Model gauge group constraint: SU(3) x SU(2) subset G4D(M)]{\boldmath The Standard Model gauge group constraint: $\SU3 \times \SU2 \subseteq G_{4D}(M)$\unboldmath}
\label{sec:constraint_G_sm}

The \emph{pseudo-GUT} constraint is a necessary condition for a model to contain the non-Abelian 
gauge group factors $G_\mathrm{SM}' = \SU3 \times \SU2$ of the SM at each step of the construction 
of the gauge embedding matrix $M$. However, our search focuses on MSSM-like models with 
$\SU{3}_\mathrm{C}\times\SU{2}_\mathrm{L}\times\U{1}_Y$ gauge symmetry in 4D and not on grand 
unified models like Pati-Salam or $\SU{5}$. Hence, after we have chosen the last vector $M_{k}$ of 
our orbifold model (taking the geometrical constraints into account), we can check that the model 
$M$ has a 4D gauge symmetry
\begin{equation}
G_{4D}(M) ~=~ \SU3 \times \SU2 \times G_\mathrm{hidden}\;.
\end{equation}
We denote this constraint by $\SU3 \times \SU2 \subseteq G_{4D}(M)$.

Due to the geometrical constraints, we first have to identify the last shift vector or Wilson line 
that can be chosen independently, i.e.\ which is not of order one and not equal to a previous 
vector. For example, for the \Z6-II orbifold geometry this results in the Wilson line $W_{6}$, as 
can be seen in table~\ref{tab:spec_MI}. However, note that for some other orbifold geometries like 
$\Z3\!\times\!\Z6$ $(2,2)$ all Wilson lines are fixed by the geometry, $W_i = (0^{16})$ for 
$i=1,\ldots,6$, and the second shift vector $V_2$ has to enable the $G_{4D}(M)$ constraint. The 
constraint is checked by computing the unbroken roots from the first $\E{8}$ factor and the sizes 
of their orthogonal root systems. This means that in order to contain $\SU3 \times \SU2$ at least 
two root systems, one of size six and another of size two, have to be present, where we allow for 
additional gauge group factors, also from the first $\E{8}$ factor. 

We implement the phenomenological constraints from section~\ref{sec:constraint_GUT} and 
section~\ref{sec:constraint_G_sm} into our search algorithm and apply it to the test case of 
$\Z{6}$-II orbifold models. It turns out that the probability of finding an MSSM-like model 
increases by a factor $10$ from $\tfrac{1}{10\,000\,000} = 10^{-7}$ in the case without the 
phenomenological constraints to $\tfrac{3}{2\,665\,463} \approx 10^{-6}$ in the case where the 
phenomenological constraints are applied. In addition, we use physical intuition that MSSM-like 
models are often related to a vanishing Wilson line~\cite{Lebedev:2008un} and perform a second 
search where we set $W_5 = (0^{16})$ by hand. The results are summarized in 
table~\ref{tab:datasets} (where the corresponding dataset is called \emph{phenomenology}).

\begin{table}[t!]
\center
\renewcommand{\arraystretch}{1.2}
\begin{tabular}{|c|l||c|r|r|r|c|c|c|}
\hline
& dataset                                        &  condition       & \# models      & \# MSSM-like & \multicolumn{4}{c|}{\# inequiv. MSSM-like} \\
\hline \hline
\multirow{4}{*}{\rotatebox[origin=c]{90}{traditional}}
& \emph{fundamental Weyl}                        &                  & $10\,000\,000$ &       1      & \multicolumn{4}{c|}{1} \\
& \emph{chamber}                                 &                  &                &              & \multicolumn{4}{c|}{}  \\
\cline{2-9}
& \multirow{2}{*}{\emph{phenomenology}}          &                  &  $2\,665\,463$ &       3      &   3 & \multicolumn{3}{c|}{\multirow{2}{*}{130}}  \\
\cline{3-6}
&                                                & $W_{5}=(0^{16})$ &  $2\,551\,272$ &     509      & 129 & \multicolumn{3}{c|}{}  \\
\hline
\hline
\multirow{4}{*}{\rotatebox[origin=c]{90}{contrast patterns\ \ \ \ }}
& \multirow{2}{*}{\emph{hidden $E_8$}}           &                  &  $2\,543\,415$ &      12      &  11 & \multirow{2}{*}{136} & \multirow{5}{*}{415} & \multirow{7}{*}{468}  \\
\cline{3-6}
&                                                & $W_{5}=(0^{16})$ &  $2\,609\,872$ &     863      & 135 & & & \\
\cline{2-7}
& \multirow{3}{*}{\emph{dynamic hidden $E_8$}}   &                  &  $1\,876\,273$ &  3\,299      & 245 & \multirow{3}{*}{395} & & \\
\cline{3-6}
&                                                & $W_{5}=(0^{16})$ &  $1\,231\,608$ &  8\,455      & 321 & & & \\
\cline{3-6}
&                                                & $W_{3}=(0^{16})$ &     $378\,604$ &       7      &   2 & & & \\
\cline{2-8}
& \multirow{2}{*}{\emph{U-sector}}             &                  &  $4\,793\,146$ &  4\,953      & 357 & \multicolumn{2}{l|}{\multirow{2}{*}{459}} &  \\
\cline{3-6}
&                                                & $W_{5}=(0^{16})$ &  $3\,046\,262$ & 17\,406      & 358 & \multicolumn{2}{c|}{} & \\
\hline
\end{tabular}
\caption{Datasets created in the \Z6-II orbifold landscape. Each dataset incorporates all of the 
constraints, indicated by the names in the first column, of the datasets above. The 
\emph{fundamental Weyl chamber} dataset utilizes the Weyl symmetry, see 
section~\ref{sec:WeylSymmetry}, while the \emph{phenomenology} dataset makes additionally use of 
the constraints developed in section~\ref{sec:pheno}. In the next rows, we apply our contrast 
patterns: first, we demand $N_{\mathrm{sr}}^{(2)} \geq 6$ from section~\ref{sec:hiddenE8} and 
obtain the \emph{hidden $E_8$} dataset. Note that in the dynamic search we modify this constraint 
as explained in section~\ref{sec:dynamicE8} to $N_{\mathrm{sr}}^{(2)} \geq X$ for 
$X \in \{8, 10, 12, \ldots, 86\}$ and obtain the \emph{dynamic hidden $E_8$} dataset. Here, the 
case $X = 6$ was disregarded since it was already sampled in the \emph{hidden $E_8$} dataset. 
Finally, the \emph{U-sector} dataset was created using the \emph{U-sector} contrast pattern from 
section~\ref{sec:u_sector} in addition. Note that we also made use of the additional 
conditions $W_5=(0^{16})$ or $W_3=W_4=(0^{16})$, where $W_5=(0^{16})$ is known to be beneficial for 
finding MSSM-like models, see ref.~\cite{Lebedev:2008un}.}
\label{tab:datasets}
\end{table}

\section[Contrast patterns for Z6-II orbifolds]{\boldmath Contrast patterns for $\Z{6}$-II orbifolds\unboldmath}
\label{sec:contrast_pattern}

In the previous section, we discussed phenomenological constraints that can be checked easily during 
the search for MSSM-like orbifold models. Importantly, these conditions are absolutely necessary 
for a model to be MSSM-like (but not sufficient). Now, we want to extend this procedure to include 
additional constraints (so-called contrast patterns) for MSSM-like models by exploiting techniques 
from data mining. These new constraints will be determined by a statistical approach. Hence, 
demanding them can potentially rule out a few MSSM-like models. In other words, the new constraints 
are not necessarily satisfied for all MSSM-like models but they significantly enhance the 
probability for a given model to be MSSM-like. In this way, we will constrain the heterotic 
orbifold landscape further to the areas of MSSM-like models. Some of these areas are hardly 
accessible by the conventional search algorithm but easy to access due to the significantly 
enhanced probability given the additional constraints from the contrast patterns.

A contrast pattern $c$ can be defined as a pattern whose supports differ significantly among the 
datasets under contrast~\cite{Dong:2012:CDM:2381030}. Here, the support is defined as
\begin{align}\label{eq:support}
\mathrm{supp}(c,D) ~=~ \dfrac{|\{M \in D \ | \ M ~\mathrm{ satisfies }~ c\} |}{|D|}\;,
\end{align}
where $D$ is a set of data points, i.e.\ orbifold models, and $c$ is a set of certain 
constraints that have to be fulfilled. In our case, we have two datasets that are under contrast: 
$D_{\text{MSSM-like}}$ and $D_{\text{\cancel{MSSM}-like}}$, which is the set of MSSM-like
models and the complementary set, respectively. In other words, we are searching for constraints 
$c$ that are satisfied for nearly all MSSM-like models while they are violated by a huge fraction 
of \cancel{MSSM}-like models. In the ideal case, we can identify contrast patterns $c$ with 
$\mathrm{supp}(c,D_{\text{MSSM-like}}) = 1$ and $\mathrm{supp}(c,D_{\text{\cancel{MSSM}-like}}) = 0$. 
This can be formalized by defining the growth rate
\begin{align}\label{eq:growth_rate}
\mathrm{gr}(c,D_{\text{MSSM-like}},D_{\text{\cancel{MSSM}-like}}) ~=~ \dfrac{\mathrm{supp}(c,D_{\text{MSSM-like}})}{\mathrm{supp}(c,D_{\text{\cancel{MSSM}-like}})}\;,
\end{align}
which has to be maximized. In the following, we will often just write $\mathrm{gr}(c)$ if the 
datasets $D_{\text{MSSM-like}}$ and $D_{\text{\cancel{MSSM}-like}}$ are clear from the context. 

To understand the growth rate better and get some intuition for its value, we rewrite it in terms 
of the probability $\hat{p}$. Here, the hat indicates that we estimate the probability by the 
sample proportion $\hat{p}(Y)=\nicefrac{N_{Y}}{N}$, where $Y\in\{$MSSM-like, \cancel{MSSM}-like$\}$ 
and the total sample size is given by $N=N_{\text{MSSM-like}} + N_{\text{\cancel{MSSM}-like}}$. 
Then, one finds
\begin{align}\label{eq:gr}
\mathrm{gr(c)}=\dfrac{\hat{p}^{c}(\text{MSSM-like})}{\hat{p}(\text{MSSM-like})}\dfrac{\hat{p}(\text{\cancel{MSSM}-like})}{\hat{p}^{c}(\text{\cancel{MSSM}-like})}\;,
\end{align}
where $\hat{p}^c(Y)=\nicefrac{N^c_Y}{N^c}$ with $N^c_Y=|\{M \in D_Y \ | \ M~\mathrm{ satisfies }~c\}|$ 
is the probability of a model being $Y=$ MSSM-like or $Y=$ \cancel{MSSM}-like given the 
constraints $c$ and $\hat{p}(Y)$ is the corresponding probability without imposing the 
constraints $c$. Then, one can solve eq.~\eqref{eq:gr} for $\hat{p}^{c}(\text{MSSM-like})$ as a 
function of $\hat{p}(\text{MSSM-like})$ as follows
\begin{align}
\hat{p}^{c}(\text{MSSM-like}) ~=~ \dfrac{\mathrm{gr(c)}\, \hat{p}(\text{MSSM-like})}{1+(\mathrm{gr(c)} -1)\hat{p}(\text{MSSM-like})}\;.
\end{align}
Here, one can observe several cases:
\begin{subequations}
\begin{align}
\hat{p}(\text{MSSM-like}) ~\ll~ 1      \ &:\ \hat{p}^{c}(\text{MSSM-like}) ~=~ \text{gr(c)}\hat{p}(\text{MSSM-like}) + \mathcal{O}\!\left(\hat{p}(\text{MSSM-like})^{2} \right)\;, \label{eq:Taylor}\\
\text{gr(c)} ~=~ 1                     \ &:\ \hat{p}^{c}(\text{MSSM-like}) ~=~ \hat{p}(\text{MSSM-like})\;, \\
\mathrm{gr(c)} ~=~ 0                   \ &:\ \hat{p}^{c}(\text{MSSM-like}) ~=~ 0\;, \\
\mathrm{gr(c)} ~\rightarrow~ \infty    \ &:\ \hat{p}^{c}(\text{MSSM-like}) ~\rightarrow~ 1\;,
\end{align}
\end{subequations}
where the Taylor expansion in eq.~\eqref{eq:Taylor} converges for 
$\hat{p}(\text{MSSM-like})<\tfrac{1}{|\mathrm{gr}(c)-1|}$. Now, eq.~\eqref{eq:Taylor} can be interpreted 
easily: For $\mathrm{gr}(c) < 1$ we have a negative effect on our favored class of MSSM-like 
models. For $\mathrm{gr}(c) = 1$ the effects on both classes cancel each other and for 
$\mathrm{gr}(c) > 1$ we have a positive effect, i.e.\ a higher probability to find MSSM-like models 
in the subspace defined by the contrast patterns $c$.

However, before we can start to search for contrast patterns $c$, we have to define some (physical) 
quantities that possibly can lead to such patterns. This is known as feature engineering as 
explained in the next section.

\subsection{Feature engineering}
\label{sec:Feature}

In this section, we will define (physical) quantities for a given orbifold model $M$. In the 
context of data mining, we will call such quantities \emph{features} and their construction is 
called \emph{feature engineering}. In general terms, feature engineering denotes the process of 
computing useful quantities from the raw data. For example, neural networks generate features in 
each hidden layer on their own during training. However, this is one of the biggest open problems 
of neural networks: it is in general very difficult to extract any meaning of the features that a 
neural network has learned on its own -- these features hardly yield any knowledge gain. 
Alternatively, in this paper we will use physical intuition and knowledge of the system to create 
features. Then, after visualizing some of these features, we can use machine learning techniques 
(i.e.\ a decision tree) to quantify if our educated guess for a certain feature, or combinations of 
multiple features, leads to a correlation between these features and the property of $M$ being 
MSSM-like or not. If such a correlation exists (i.e.\ if $\mathrm{gr}(c) > 1$), we have identified a 
promising contrast pattern $c$. Moreover, if we can check this pattern $c$ easily in the search 
algorithm displayed in figure~\ref{fig:successive_create}, we can utilize it to reduce the search 
space in the heterotic orbifold landscape to areas where MSSM-like orbifold models accumulate. The 
advantage of this approach is obvious: by construction we have a straightforward physical 
interpretation of our features.

\begin{figure}[ht!]
\includegraphics[width=12cm]{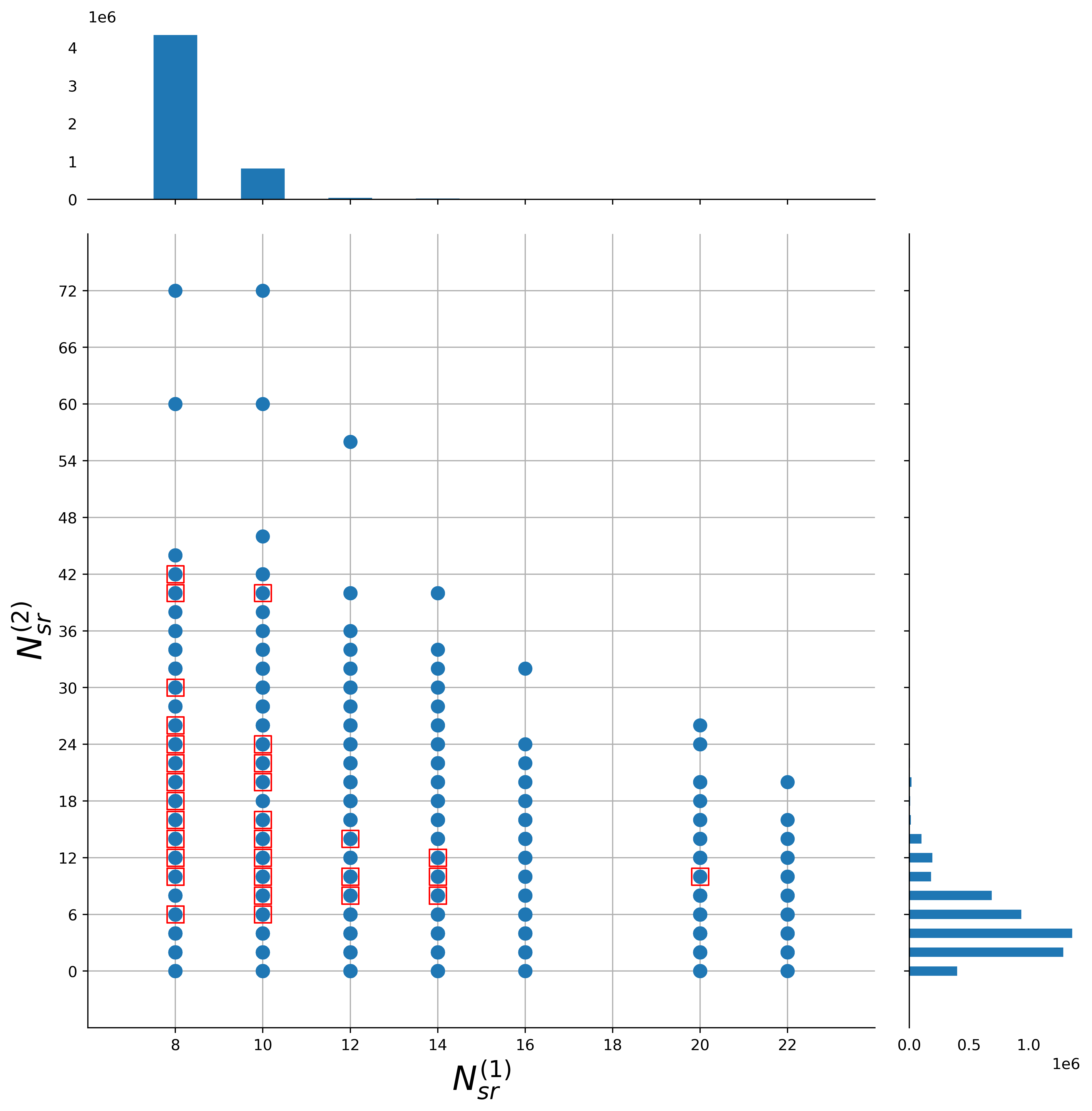}
\centering
\caption{Jointplot (see ref.~\cite{michael_waskom_2017_883859}) of $N_{\mathrm{sr}}^{(1)}$ and 
$N_{\mathrm{sr}}^{(2)}$ for the complete \emph{phenomenology} dataset of $\Z{6}$-II orbifold 
models, see table~\ref{tab:datasets}. Due to the constraint $\SU3 \times \SU2 \subseteq G_{4D}(M)$ 
from section~\ref{sec:constraint_G_sm} we have a bound $N_{\mathrm{sr}}^{(1)} \geq 8$ in this 
dataset. The central part shows a scatter plot, where \cancel{MSSM}-like and MSSM-like orbifold 
models are marked by blue circles and red boxes, respectively. Above and on the right hand side of 
the scatter plot we give the histogram on the number of models for given values of 
$N_{\mathrm{sr}}^{(1)}$ and $N_{\mathrm{sr}}^{(2)}$, respectively. Note that the number of 
MSSM-like models is negligible compared to the \cancel{MSSM}-like models in this dataset. 
Hence, the histograms visualize the frequency of occurrence for \cancel{MSSM}-like models only.}
\label{fig:jointplot}
\end{figure}

However, in general it is very difficult to identify useful features. In our case, we have tried 
many concepts, for instance, local GUTs, the breaking patterns of each shift vector and each Wilson 
line, the breaking patterns of certain combinations thereof, the number of non-Abelian gauge group 
factors in 4D, and many more. We know from section~\ref{sec:setup_random_search} that the 4D gauge 
group has a great impact and it can be checked easily at every step of the production of a model. 
Therefore, there is hope that additional features can be found from the 4D gauge group. When 
attempting to identify a promising feature, data visualization can be very useful: In 
figure~\ref{fig:jointplot}, we plot the number of unbroken roots $N_{\mathrm{sr}}^{(\alpha)}$ from 
each $\E{8}$ factor in a scatter plot against each other using the \emph{phenomenology} dataset
created in section~\ref{sec:pheno}. In addition, the respective histograms for the number of 
orbifold models with a certain number of unbroken roots are displayed in 
figure~\ref{fig:jointplot} for each axis (i.e.\ for each $\E{8}$ factor).
To see if the feature $N_{\mathrm{sr}}^{(\alpha)}$ has the potential to be useful as a contrast 
pattern, we have to pay attention to two aspects of this plot. First, we have to identify areas in 
this plot where no MSSM-like orbifold model is present. Second, it is important that such an area 
is not only qualitatively separated from MSSM-like orbifold models but also quantitatively 
interesting, i.e.\ highly populated with \cancel{MSSM}-like orbifold models. This aspect can be 
read off from the respective histogram in figure~\ref{fig:jointplot}. By doing so, we identify the 
area $N_{\mathrm{sr}}^{(2)} < 6$ that does not contain any MSSM-like $\Z{6}$-II orbifold model 
but is fairly high populated with \cancel{MSSM}-like models in our \emph{phenomenology} dataset. 
Hence, the condition $N_{\mathrm{sr}}^{(2)} < 6$ is our first promising candidate for a contrast 
pattern and we expect to get a strong reduction of the $\Z{6}$-II orbifold landscape by 
excluding this area form the search. In contrast, an example of an area that consists of 
\cancel{MSSM}-like $\Z{6}$-II models only but is irrelevant due to the small number of models 
is given by $N_{\mathrm{sr}}^{(2)} > 42$ or $N_{\mathrm{sr}}^{(1)} > 20$: the respective histograms 
(in figure~\ref{fig:jointplot}, the vertical histogram on the right-hand side for 
$N_{\mathrm{sr}}^{(2)} > 42$ and the horizontal histogram above the scatter plot for 
$N_{\mathrm{sr}}^{(1)} > 20$) show that the number of $\Z{6}$-II orbifold models in these 
regions is very small.

In addition to the above features, we will also use the numbers of orbifold-invariant bulk matter
fields as additional features. They are computed similar to the number of unbroken roots of the 
gauge group in eq.~\eqref{eq:number_surviving_roots}, with the difference of an additional 
displacement from the geometrical twist vector $v$, i.e.\ at each step $n$ of our search algorithm 
displayed in figure~\ref{fig:successive_create} we compute
\begin{align}\label{eq:number_U_sector}
& N_{\mathrm{U}_{a}}^{(\alpha)}(M) ~=~ \sum_{p \,\in\, \Phi_{\E{8}}} \prod\limits_{k\,=\,1}^{n} \delta\left( p \cdot M_{k}^{(\alpha)} - \Theta(2-k)\, q_{(a)} \cdot v_{(k)} \right)\;,
\end{align}
for $\alpha=1,2$ and $a=1,2,3$. Note that the term $q_{(a)} \cdot v_{(k)}$ in 
eq.~\eqref{eq:number_U_sector} is turned off for the Wilson lines $M_{k}^{(\alpha)}$, 
$k=3,\ldots,8$, using
\begin{align}
\Theta(x) ~=~
 \begin{cases} 0 \quad\mathrm{if} & x < 0\\
               1 \quad\mathrm{if} & x \geq 0\;.
 \end{cases}
\end{align}
Furthermore, the vectors $q_{(1)}=(0,-1,0,0)$, $q_{(2)}=(0,0,-1,0)$ and $q_{(3)}=(0,0,0,-1)$ give 
rise to the three untwisted sectors $\mathrm{U}_a$, $a=1,2,3$, and correspond to the three 
directions of the internal vector-boson index of the ten-dimensional $\E{8}\times\E{8}$ gauge 
bosons, respectively. Finally, the twist vectors in eq.~\eqref{eq:number_U_sector} are given by 
$v_{(1)}=(0,\tfrac{1}{6},\tfrac{1}{3},-\tfrac{1}{2})$ and $v_{(2)}=(0^{4})$ is not present for the 
\Z6-II orbifold geometry. Due to the $\delta$-condition in eq.~\eqref{eq:number_U_sector} all 
features $N_{\mathrm{U}_{a}}^{(\alpha)}(M)$ and $N_{\mathrm{sr}}^{(\alpha)}(M)$ are independent.

\begin{table}[ht!]
\center
\renewcommand{\arraystretch}{1.4}
\begin{tabular}{|l|cccc|cccc|}
\hline
 & \multicolumn{4}{c|}{first $\E{8}$} & \multicolumn{4}{c|}{hidden $\E{8}$} \\ 
 \hline \hline
feature & $N_{\mathrm{sr}}^{(1)}$ & $N_{\mathrm{U}_{1}}^{(1)}$ & $N_{\mathrm{U}_{2}}^{(1)}$ & $N_{\mathrm{U}_{3}}^{(1)}$ & $N_{\mathrm{sr}}^{(2)}$ & $N_{\mathrm{U}_{1}}^{(2)}$ & $N_{\mathrm{U}_{2}}^{(2)}$ & $N_{\mathrm{U}_{3}}^{(2)}$ \\\hline
\end{tabular}
\caption{Overview table of the features that we use for contrast mining. Each feature is evaluated 
for all orbifold models $M$ of the dataset under investigation.}
\label{tab:features}
\end{table}

\subsection{Decision tree and \emph{false negatives}}
\label{sec:decisiontree}

In this section, we will use a decision tree~\cite{scikit-learn} in order to identify those 
features from table~\ref{tab:features} that correlate with the property of a model being MSSM-like 
or not. If such a correlation exists, the corresponding feature can be used as contrast pattern in 
our search algorithm for MSSM-like orbifold models. A decision tree belongs to the class of 
supervised machine learning and can be used for the purpose of classification or regression. In our 
case, we want to classify whether a given orbifold model is MSSM-like or not using simple 
true-or-false decisions on the features listed in table~\ref{tab:features}. Then, by analyzing the 
decisions made inside of the decision tree, we can identify those features that lead to successful 
contrast patterns.

In more detail, our decision tree is a function from some of the features listed in 
table~\ref{tab:features} to the classification value denoted by $Y$, i.e.\ it is of the form
\begin{equation}\label{eq:decisiontree}
\left(N_{\mathrm{sr}}^{(1)}(M), N_{\mathrm{U}_{1}}^{(1)}(M),N_{\mathrm{U}_{2}}^{(1)}(M),N_{\mathrm{U}_{3}}^{(1)}(M),\ldots, \right) ~\mapsto~ Y_\mathrm{predicted}(M)\;,
\end{equation}
where $Y_\mathrm{predicted}(M) \in \{\text{MSSM-like}, \text{\cancel{MSSM}-like}\}$, and it can be 
applied to all orbifold models $M$. Note that we can compute for each orbifold model $M$ both, the 
features and the correct classification value 
$Y_\mathrm{correct}(M) \in \{\text{MSSM-like}, \text{\cancel{MSSM}-like}\}$ using the 
\texttt{orbifolder}. Yet, the benefit of using a decision tree is given by the possibility to uncover 
unknown correlations between our features and the property $Y_\mathrm{correct}(M)$ of an orbifold 
model $M$ to be MSSM-like or not. Furthermore, it is by no means guaranteed that a function like 
eq.~\eqref{eq:decisiontree} exists. We will only know about its existence after we have trained and 
tested our decision tree.

In a first step, the decision tree has to be trained, i.e.\ the algorithm tries to learn the 
function~\eqref{eq:decisiontree}. To do so, it needs a training set, i.e.\ a list of orbifold 
models, where for each orbifold model $M_i$ we have computed the values of our features and the 
correct classification values $Y_\mathrm{correct}(M_i)$ using the \texttt{orbifolder}. More 
explicitly, the training set reads
\begin{eqnarray}
\text{training set} &=& \{\{N_{\mathrm{sr}}^{(1)}(M_1), N_{\mathrm{U}_{1}}^{(1)}(M_1),N_{\mathrm{U}_{2}}^{(1)}(M_1),N_{\mathrm{U}_{3}}^{(1)}(M_1),\ldots, Y_\mathrm{correct}(M_1)\}, \nonumber\\
                    & & \,\,\{N_{\mathrm{sr}}^{(1)}(M_2), N_{\mathrm{U}_{1}}^{(1)}(M_2),N_{\mathrm{U}_{2}}^{(1)}(M_2),N_{\mathrm{U}_{3}}^{(1)}(M_2),\ldots, Y_\mathrm{correct}(M_2)\}, \ldots\}\;,
\end{eqnarray}
and during training the decision tree tries to adjust the function eq.~\eqref{eq:decisiontree} 
such that $Y_\mathrm{predicted}(M) = Y_\mathrm{correct}(M)$ for all models $M$ from the training 
set. After training, we can evaluate the trained decision tree~\eqref{eq:decisiontree} on the 
so-called validation set 
\begin{eqnarray}
\text{validation set} &=& \{\{N_{\mathrm{sr}}^{(1)}(P_1), N_{\mathrm{U}_{1}}^{(1)}(P_1),N_{\mathrm{U}_{2}}^{(1)}(P_1),N_{U\mathrm{U}_{3}}^{(1)}(P_1),\ldots, Y_\mathrm{correct}(P_1) )\}, \nonumber\\
                    & & \,\,\{N_{\mathrm{sr}}^{(1)}(P_2), N_{\mathrm{U}_{1}}^{(1)}(P_2),N_{\mathrm{U}_{2}}^{(1)}(P_2),N_{U\mathrm{U}_{3}}^{(1)}(P_2),\ldots, Y_\mathrm{correct}(P_2)\}, \ldots\}\;,
\end{eqnarray}
containing the features of some other orbifold models $P_i$. Then, we can compare the results 
$Y_\mathrm{predicted}(P_i)$ of the decision tree~\eqref{eq:decisiontree} to the correct values 
$Y_\mathrm{correct}(P_i)$ obtained from the \texttt{orbifolder}. In this way, we can check 
whether our decision tree was able to identify the function eq.~\eqref{eq:decisiontree} between our 
features and the property of a model being MSSM-like or not. 

In practice, a decision tree 
will not be trained perfectly. First of all, it is possible that there is no exact functional 
dependency of the form eq.~\eqref{eq:decisiontree}. Furthermore, even in the case when such a 
functional dependency would exist in principle, the decision tree might be unable to learn it, 
possibly because the training set was too small or imbalanced. In general, we can distinguish 
between two types of errors, i.e.\ cases where $Y_\mathrm{correct}(M) \neq Y_\mathrm{predicted}(M)$. 
They are called:
\begin{itemize}
\item \emph{false positives}: $Y_\mathrm{correct}(M)=$ \cancel{MSSM}-like but $Y_\mathrm{predicted}(M)=$ MSSM-like
\item \emph{false negatives}: $Y_\mathrm{correct}(M)=$ MSSM-like but $Y_\mathrm{predicted}(M)=$ \cancel{MSSM}-like
\end{itemize}
Every classification process tries to minimize the number of false predictions. However, at a 
certain level it always comes to a trade-off between \emph{false positives} and \emph{false 
negatives} and we have to decide whether we want to suppress one of them for the drawback of 
raising the other one. In our case, a \emph{false positive} classification by the decision tree is 
not a big problem since we can simply check each of these orbifold models afterwards explicitly 
using the \texttt{orbifolder}. However, in the case of a \emph{false negative} classification the 
consequences are that we will loose an MSSM-like orbifold model. Since MSSM-like orbifold models 
are far too valuable to us, we want to minimize the number of \emph{false negative} cases by all 
means, while we want to keep the number of \emph{false positives} as low as possible. Therefore, we 
introduce a loss matrix $L$, which informs the machine learning algorithm about the different 
importance of certain models~\cite{Bishop:2006:PRM:1162264}. We choose a loss matrix
\begin{align}
L =
\left(
\begin{array}{cc}
0 & 10^{6} \\
1 &  0
\end{array}
\right)~,
\end{align}
where the two rows correspond to the correct value $Y_\mathrm{correct}(M)$ being either MSSM-like 
or not, and the two columns correspond to the predicted values $Y_\mathrm{predicted}(M)$ being 
either MSSM-like or not. Then, $L_{12}$ corresponds to the \emph{false negative} cases of an 
MSSM-like orbifold model $M$ that has been classified by the decision tree to be 
$Y_\mathrm{predicted}(M)=$ \cancel{MSSM}-like. As this is very undesirable, the system is punished 
with a large loss value $L_{12}=10^6$. As discussed before, the other possible error of a 
\emph{false positive} classification is not so severe. Hence, we set $L_{21}=1$. This will guide 
the decision tree algorithm towards suppressing the \emph{false negative} cases such that we do not 
miss any MSSM-like orbifold models.

For later convenience, we will quantify the quality of the predictions by the recall. It is defined 
as the number of correct predictions of the MSSM-like class divided by the total number of 
MSSM-like orbifold models. Hence, if the number of \emph{false negatives} for all MSSM-like 
orbifold models $P_i$ is zero, the recall is $1.00$ on the validation set and all MSSM-like 
orbifold models are assigned with the correct value $Y=$ MSSM-like.

In the following, we apply decision trees to our features in order to extract promising contrast 
patters.

\subsubsection[The hidden E8 contrast pattern]{\boldmath The \emph{hidden $E_8$} contrast pattern\unboldmath}
\label{sec:hiddenE8}

As a first step, we have to define our datasets. The training and validation set are created by a 
random split (using a validation size of 33\%) of the \emph{phenomenology} dataset from 
table~\ref{tab:datasets} (based on the phenomenological constraints from 
section~\ref{sec:constraint_GUT} and section~\ref{sec:constraint_G_sm}). However, we add a small 
modification to the dataset to avoid \emph{data leakage}. Data leakage refers to the mistake to 
inform the machine learning algorithm about data from the validation set during training. In this 
case, the machine learning algorithm might overfit on some of the data from the validation set even 
though the data was divided into training and validation set. As the performance of a machine 
learning model on the validation set is a measure for its ability to generalize to unseen data, 
this mistake has to be avoided. In our case this could happen if, for example, there exists an 
MSSM-like model that completely dominates all MSSM-like models with all of its equivalent copies. 
Then, this model would appear most likely in both, training and validation set. Hence, the machine 
learning algorithm would see this model during training. Moreover, the same model would dominate 
the results on the validation set and pretend that the learned predictions generalize to generic 
MSSM-like models. Therefore, to avoid data leakage we only use inequivalent MSSM-like models. 
Nevertheless, for the \cancel{MSSM}-like models we keep the equivalent models, since the frequency 
of occurrence gives us a notion of the size of the area that a certain split in the decision tree 
excludes. In more detail, we have to perform our search for MSSM-like orbifold models in the space 
$\Z{}^{128}$ of gauge embedding matrices $\{M_i\}$, see eq.~\eqref{eq:ExpansionInDualBasis}, even 
though we are interested in the space of physically inequivalent models $\{\mathrm{model}(M_j)\}$, 
or more precisely, in the space of inequivalent massless particle spectra $\{\mathrm{spectrum}(M_k)\}$. 
Now, we defined features that directly depend on $\mathrm{spectrum}(M)$, not on $d\in\Z{}^{128}$ 
and our decision tree performs its splits based on these features. Consequently, a certain split in 
the decision tree will exclude all points $d\in\Z{}^{128}$ that give rise to the same excluded 
features. In this way, a small restriction in feature space gives rise to a huge effect in the 
space $\Z{}^{128}$ of gauge embedding matrices. Moreover, by not restricting the feature space too 
much, we leave enough room to discover new MSSM-like models, also in unexpected areas of the 
landscape. In contrast, if we had used only inequivalent \cancel{MSSM}-like models for the 
training of our decision tree, the decision tree would not care to exclude a single model even 
though it might actually correspond to a huge area in $\Z{}^{128}$. At the end, we want to enhance 
the search algorithm. Hence, it is better to exclude a few models with extraordinary high frequency 
of occurrence than multiple models with very low one.

\begin{figure}[t!]
\includegraphics[width=12cm]{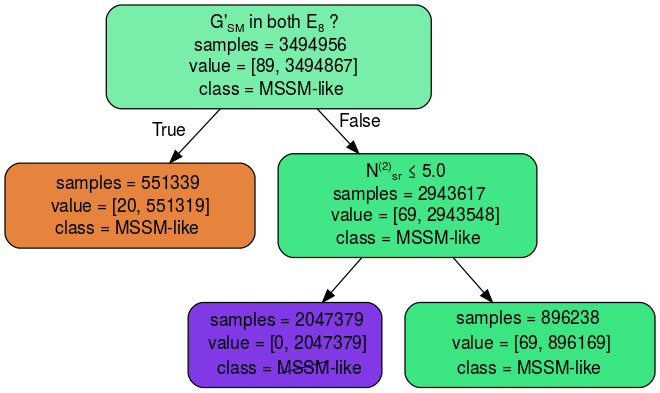}
\centering
\caption{Decision tree on the \emph{phenomenology} dataset from table~\ref{tab:datasets}, evaluated 
on the training set. We can extract the contrast pattern $N_{\mathrm{sr}}^{(2)} \leq 5$ for 
\cancel{MSSM}-like models from this tree, which we reformulate into a positive condition 
$N_{\mathrm{sr}}^{(2)} \geq 6$ for MSSM-like models. Let us explain the details at the example of 
the uppermost node. This node contains: (i) the condition \emph{$G'_{SM}$ in both $E_8$?} that has 
to be evaluated, (ii) \emph{samples=3\,494\,956} gives the total number of models in this node, 
(iii) \emph{value = $[89, 3\,494\,867]$} gives the number of MSSM-like and \cancel{MSSM}-like 
models in this node, respectively, and, finally, (iv) \emph{class = MSSM-like} is the prediction 
for all models in this intermediate node. The final prediction for the models is given by the leaf 
nodes.
}
\label{fig:tree_staticE8}
\end{figure}

Now, we train the decision tree on the training set. Here, we tune the hyperparameters such that 
we get a recall value for MSSM-like models of $1.00$ on the validation set. This is due to the fact 
that we want to find contrast patterns that are satisfied by all MSSM-like models contained in the 
validation set. During training, the decision tree identifies areas in feature space and 
assigns the two classes $\{\text{MSSM-like}, \text{\cancel{MSSM}-like}\}$ to them, using the data 
from the training set. However, we want the decision tree to assign the class \cancel{MSSM}-like 
only to those areas that are also highly populated with \cancel{MSSM}-like models. In this way, we 
can be sure that the probability for an MSSM-like model is extremely small in these areas of 
\cancel{MSSM}-like models. This can be achieved using a technique called pruning. Consequently, the 
complexity of the decision tree is reduced to a minimum and we keep the possibility to find 
MSSM-like models in rather unexpected areas of the landscape. 

The resulting decision tree is displayed in figure~\ref{fig:tree_staticE8}. We see that the 
intuition given by the scatter plot in figure~\ref{fig:jointplot} manifests in a lower bound on the 
number of unbroken roots from the hidden $\E{8}$ factor: From the second node in the second line of 
the decision tree we can extract the condition
\begin{equation}\label{eq:FirstCP}
c_\text{hidden \E{8}} ~=~\Big\{N_{\mathrm{sr}}^{(2)}(M) ~\geq~ 6\Big\} \;,
\end{equation} 
for a model $M$ to have a high probability to be MSSM-like. We call this condition the \emph{hidden 
$E_8$} contrast pattern.  In principle, our decision tree contains additional splits. However, we 
want to stay conservative with our search and avoid too enthusiastic splits. Hence, we stay with 
the first and most important split for now. Based on the training set form our \emph{phenomenology} 
dataset we can estimate the growth rate of the \emph{hidden $E_8$} contrast pattern to be
\begin{equation}\label{eq:estimatedgrowthrate}
\mathrm{gr}\left(c_{\text{hidden \E{8}}},D_{\text{MSSM-like}},D_{\text{\cancel{MSSM}-like}}\right) ~=~ \dfrac{1}{\tfrac{896\,169+551\,319}{3\,494\,867}} ~\approx~ 2.4 ~>~ 1\;,
\end{equation}
using eq.~\eqref{eq:growth_rate} with $\mathrm{supp}(c_{\text{hidden \E{8}}},D_{\text{MSSM-like}})=1$ for our 
\emph{phenomenology} dataset and the numbers for $\mathrm{supp}(c_{\text{hidden \E{8}}},D_{\text{\cancel{MSSM}-like}})$ 
can be read off from figure~\ref{fig:tree_staticE8}. In other words, we can modify our search 
algorithm such that we would have avoided $2\,047\,379$ \cancel{MSSM}-like models in the training 
set. Hence, the contrast pattern~\eqref{eq:FirstCP} allows us to exclude a huge area in the 
$\Z{6}$-II orbifold landscape which statistically does not lead to MSSM-like models. Instead, we 
can invest the gained computing time to search in areas where the probability of a model to be 
MSSM-like is significantly increased. 

We implement the \emph{hidden $E_8$} contrast pattern into our search algorithm displayed in 
figure~\ref{fig:successive_create} and perform an intensive search using in addition different 
constraints on the Wilson lines: First, we allow all Wilson lines to be non-trivial and then, 
motivated by ref.~\cite{Lebedev:2008un}, we turn off $W_5$ by hand. The resulting dataset is 
called \emph{hidden $E_8$} and summarized in table~\ref{tab:datasets}. One observes an increase of 
the probability to find an MSSM-like model from 
\begin{subequations}
\begin{eqnarray}
\hat{p}^\text{phenomenology}(\text{MSSM-like})  & = & \frac{512}{5\,216\,735} ~\approx~ 10^{-4}\quad\mathrm{to}  \\
\hat{p}^{\text{hidden \E{8}}}(\text{MSSM-like}) & = & \frac{875}{5\,153\,287} ~\approx~ 2 \cdot 10^{-4}\;,
\end{eqnarray}
\end{subequations}
which is consistent with the estimated growth rate in eq.~\eqref{eq:estimatedgrowthrate}. However, 
these are the probabilities to find any MSSM-like model and it does not need to be inequivalent 
to the already known ones. Unfortunately, the total number of inequivalent MSSM-like models did 
not satisfy our expectations: it increased from 130 inequivalent MSSM-like models in the 
\emph{phenomenology} dataset to 136 in the \emph{hidden $E_8$} dataset. In the next section, we 
will investigate the reasons for this and present a solution that will lead to many new 
inequivalent MSSM-like models.

\subsubsection[The dynamic hidden E8 contrast pattern]{\boldmath The \emph{dynamic hidden $E_8$} contrast pattern\unboldmath}
\label{sec:dynamicE8}

\begin{figure}[ht!]
\includegraphics[width=\textwidth]{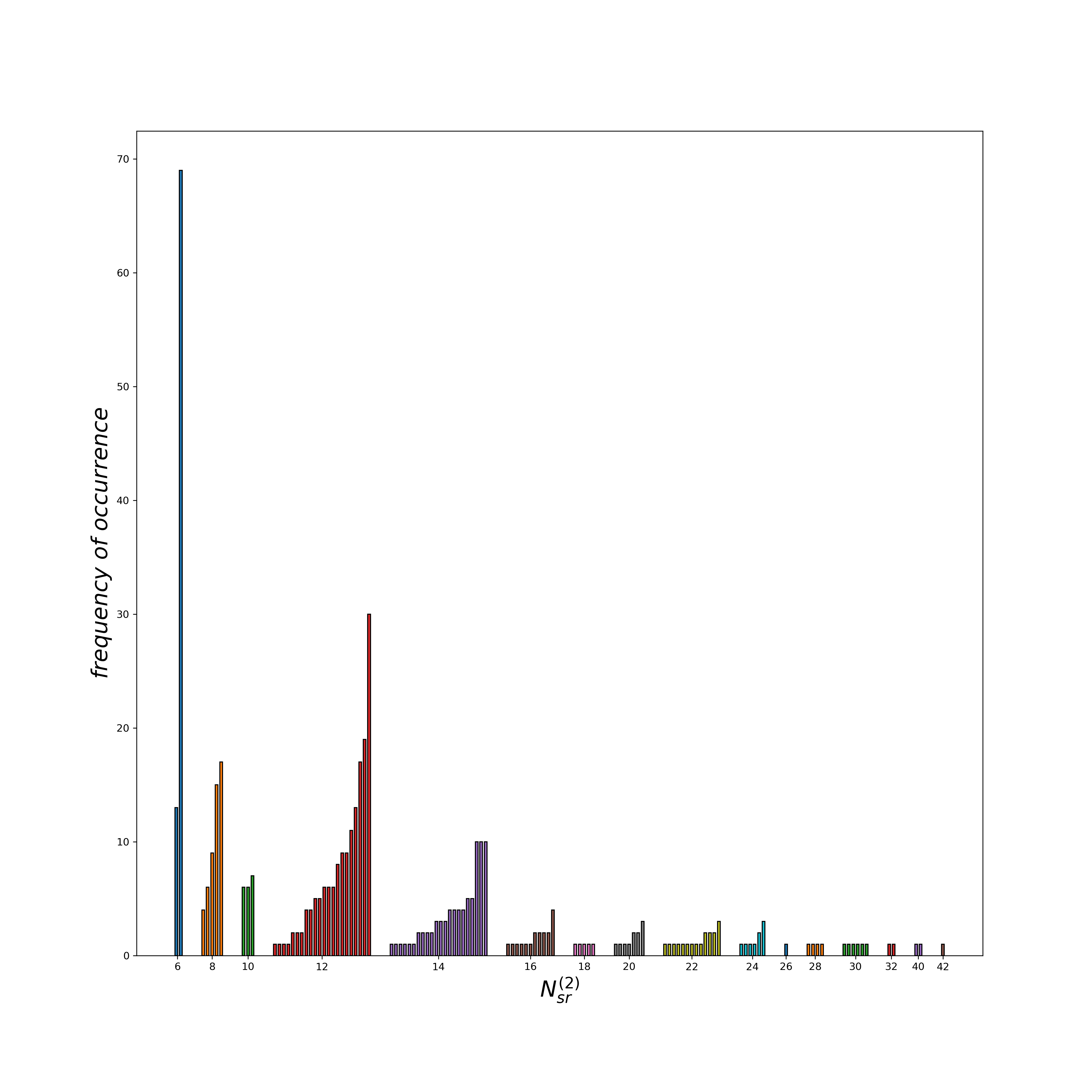}
\centering
\vspace{-2cm}
\caption{Bar chart of MSSM-like $\Z{6}$-II models $M$ found using the \emph{hidden $E_8$} 
contrast pattern $N_{\mathrm{sr}}^{(2)}(M) \geq 6$. On the horizontal axis we give 
$N^{(2)}_{\mathrm{sr}}(M)$ corresponding to the number of unbroken roots from the hidden $\E{8}$. 
The vertical axis gives the corresponding frequency of occurrence. Each bar stands for an 
inequivalent MSSM-like model $M$. The number of copies of each model is shown by the height of the 
bar. As an example, take $N^{(2)}_{\mathrm{sr}}(M) = 10$, i.e.\ the green bars: There are three 
inequivalent MSSM-like models, each represented by one bar. These inequivalent models have 
6, 6 and 7 copies in the whole dataset, respectively. Note that in this chart only those MSSM-like 
models appear that have $G_{SM}'$ only in one $\E{8}$, since the notion of hidden $\E{8}$ is 
ambiguous otherwise.}
\label{fig:inequ_Repetition_vs_SurvRoots}
\end{figure}

Next, we analyze the effect of the \emph{hidden $E_8$} contrast pattern in more detail in order to 
identify a way to improve this constraint further. To do so, we take the (equivalent) MSSM-like 
$\Z{6}$-II models from the \emph{hidden $E_8$} dataset and visualize how many MSSM-like models $M$ 
appear for various values of $N_{\mathrm{sr}}^{(2)}(M)$, see 
figure~\ref{fig:inequ_Repetition_vs_SurvRoots}.
From this chart we see that the models with small numbers of unbroken roots 
$N^{(2)}_{\mathrm{sr}}(M)\in\{6, \ldots,14\}$ are heavily oversampled in the \emph{hidden $E_8$} 
dataset, while it seems to be very difficult to construct models with $N^{(2)}_{\mathrm{sr}}(M)\geq 30$. 
Moreover, note that especially for $N^{(2)}_{\mathrm{sr}}(M)=22$ the bar chart shows a lot of 
different bars, i.e.\ there are many inequivalent MSSM-like models with 
$N^{(2)}_{\mathrm{sr}}(M)=22$. This suggests that the diversity of inequivalent MSSM-like models 
may lie in some areas of the $\Z{6}$-II orbifold landscape where models have larger hidden sector 
gauge groups. Furthermore, we investigate the change of the growth rate for higher threshold 
values $X$ of our contrast pattern $N^{(2)}_{\mathrm{sr}}(M) \geq X$ and obtain 
table~\ref{tab:growthrateOfX}. Therefore, it seems very promising to change the threshold value 
$X=6$ of the contrast pattern $N^{(2)}_{\mathrm{sr}}(M) \geq X$ into a dynamic variable $X$. We 
call this new constraint \emph{dynamic hidden $E_8$}. By applying this dynamic contrast pattern, we 
hope that the sampling among the various sizes of the hidden sector gets more balanced. 
Furthermore, we expect a boost in the number of MSSM-like models due to the increasing growth rate 
for higher values of $N^{(2)}_{\mathrm{sr}}(M)$. 

\begin{table}[t!]
\center
\renewcommand{\arraystretch}{1.5}
\begin{tabular}{|c||ccccccccccccc|}
\hline
$X$ & \makebox[0.5cm]{6}  & \makebox[0.5cm]{8} & \makebox[0.5cm]{10} & \makebox[0.5cm]{12} &\makebox[0.5cm]{\dots} & \makebox[0.5cm]{20} & \makebox[0.5cm]{22} & \makebox[0.5cm]{\dots} & \makebox[0.5cm]{30} & \makebox[0.5cm]{32} & \makebox[0.5cm]{\dots} & \makebox[0.5cm]{40} & \makebox[0.5cm]{42}\\ 
\hline
$\mathrm{gr}(N^{(2)}_{\mathrm{sr}}(M) \geq X)$ &
1 & 5 & 6 & 6 & \dots & 12 & 26 & \dots & 48 & 75 & \dots & 73  & 82 \\
\hline
\end{tabular}
\caption{Change of the growth rate for higher threshold values $X$ of our contrast pattern 
$N^{(2)}_{\mathrm{sr}}(M) \geq X$, where the reference point 
is $\mathrm{gr}(N^{(2)}_{\mathrm{sr}}(M) \geq 6)=1$, since this analysis is done in the 
\emph{hidden $E_8$} dataset.}
\label{tab:growthrateOfX}
\end{table}

We perform an intensive search based on the \emph{dynamic hidden $E_8$} contrast pattern for 
various values of the threshold $X$ and different constraints on the Wilson lines: First, we allow 
all Wilson lines to be non-trivial, then we turn off either $W_3=W_4$ or $W_5$. As a result, we 
obtain a new dataset (which we also call \emph{dynamic hidden $E_8$}), see table~\ref{tab:datasets}. 
Compared to the \emph{hidden $E_8$} dataset with 136 inequivalent MSSM-like $\Z{6}$-II models we 
now have in total 415 MSSM-like models. This is already more than in any existing $\Z{6}$-II 
search~\cite{Lebedev:2006kn,Lebedev:2008un,Nilles:2014owa,Olguin-Trejo:2018wpw}. Hence, we were 
able to significantly improve the search for inequivalent MSSM-like models in the $\Z{6}$-II 
orbifold landscape. Moreover, this search solves the puzzle of the absence of MSSM-like models in 
the case $W_3=W_4=(0^{16})$: So far, it was not possible to find any MSSM-like model if the order 3 
Wilson line is turned off, even though there is no theoretical obstruction for such a model to 
exist. Now, we have identified two MSSM-like $\Z{6}$-II models with $W_3=W_4=(0^{16})$ as can be 
seen in table~\ref{tab:datasets}. These models are equipped with a phenomenologically appealing 
$\Delta(54)$ flavor symmetry. Thus, we present these models in some detail in 
section~\ref{sec:delta54}.

\begin{figure}[t!]
\includegraphics[width=\textwidth]{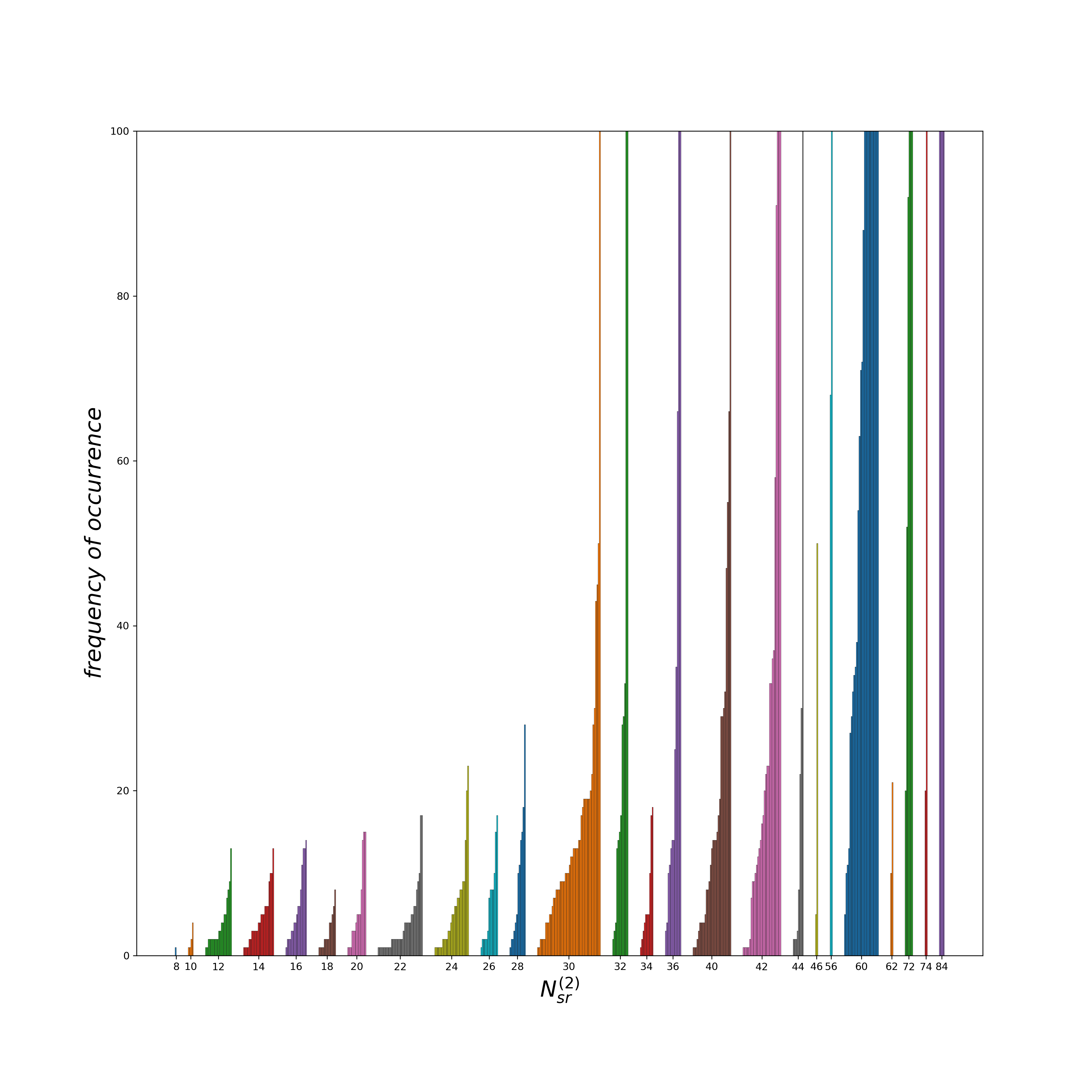}
\centering
\vspace{-2cm}
\caption{
Bar chart of MSSM-like $\Z{6}$-II models $M$ found using the \emph{dynamic hidden $E_8$} contrast 
pattern, c.f.\ figure~\ref{fig:inequ_Repetition_vs_SurvRoots}. Note that increasing the threshold 
value $X$ of the contrast pattern $N^{(2)}_{\mathrm{sr}}(M) \geq X$ leads to a deeper search in 
those areas of the $\Z{6}$-II orbifold landscape that were insufficiently sampled by the static 
search using $X=6$.}
\label{fig:inequ_Repetition_vs_SurvRoots_dyn}
\end{figure}

A few remarks are in order. It is clear that previous searches based on the traditional approach as 
well as those presented in this paper are in general not exhaustive. During any random search 
process the number of inequivalent MSSM-like models will follow a saturation 
curve~\cite{Dienes:2006ca}. Consequently, the effort for creating a new inequivalent MSSM-like 
model growth exponentially during sampling. Thus, we believe that any attempt to reach our result 
using a basic random search would take an unrealizable amount of computing time and should be 
considered only a theoretical possibility rather then an alternative approach. So, why is our new 
search strategy so successful? Astonishingly, it turns out that a huge fraction of the diversity of 
MSSM-like models lies in areas of the heterotic orbifold landscape where the hidden sector gauge 
group is large, see figure~\ref{fig:inequ_Repetition_vs_SurvRoots_dyn}. In more detail, using the 
\emph{dynamic hidden $E_8$} contrast pattern we could (i) obtain many new MSSM-like models with 
$N^{(2)}_{\mathrm{sr}}(M) = X$ for $X \in \{34, 36, 44, 46, 56, 60, 62, 72, 74, 84\}$ and (ii) 
resolve the richness of MSSM-like models for higher $X$ values, e.g. with $X\in\{30, 40, 42\}$. 
These large hidden sector gauge groups can have direct physical implications related to 
supersymmetry breaking via gaugino condensation at rather high energies~\cite{Lebedev:2006tr} and 
have to be studied in more detail.

\subsubsection[The U-sector contrast pattern]{\boldmath The \emph{U-sector} contrast pattern\unboldmath}
\label{sec:u_sector}

On the basis of our \emph{hidden $E_8$} and \emph{dynamic hidden $E_8$} datasets we want to search 
for further contrast patterns. To do so, we follow the same logic as in section~\ref{sec:hiddenE8} 
and apply a decision tree on the remaining features $N_{\mathrm{U}_{a}}^{(\alpha)}(M)$ to our new, 
combined dataset. For computational reasons we downsample our background of \cancel{MSSM}-like 
models. This means we only work with a fraction of $\sim 50\%$ of the total dataset. This is a 
valid approach since we have so much data that the actual statistics for the decision tree will not 
change in a relevant way even if the whole dataset had been given. Furthermore, for the rare and 
important MSSM-like models we keep all inequivalent MSSM-like models, as described in 
section~\ref{sec:hiddenE8}.

\begin{figure}[t!]
\includegraphics[width=12cm]{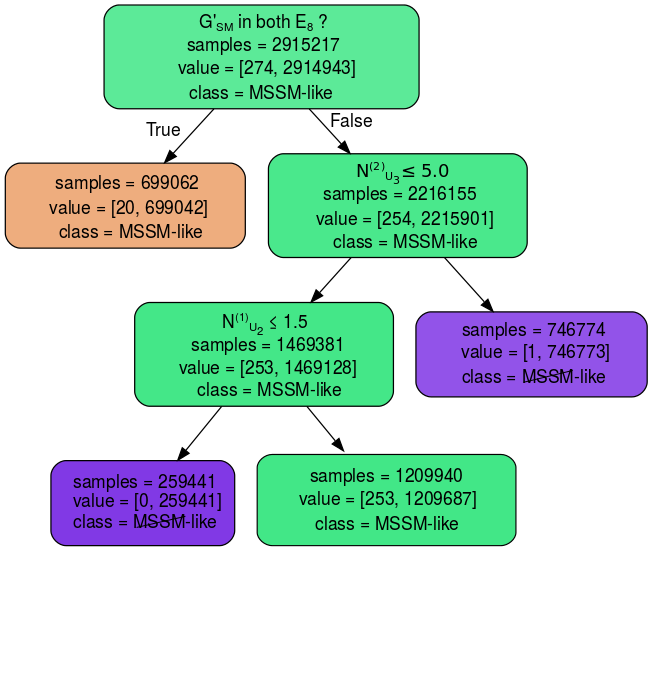}
\centering
\vspace{-2.7cm}
\caption{Decision tree on the \emph{hidden $E_8$} and \emph{dynamic hidden $E_8$} datasets from 
table~\ref{tab:datasets}, evaluated on the training set. We can extract the contrast pattern 
($N_{\mathrm{U}_2}^{(1)} \geq 2$ and $N_{\mathrm{U}_3}^{(2)} \leq 5$) for MSSM-like models from 
this tree. One observes that this tree misclassifies one MSSM-like model after the 
$N_{\mathrm{U}_3}^{(2)}$-split in order to get better performance. See also 
figure~\ref{fig:tree_staticE8} for further details.}
\label{fig:tree_dyn}
\end{figure}

Then, the decision tree is trained with the same aim as before to classify all MSSM-like models 
correctly in both, training and validation set. However, it turns out that this is rather difficult 
due to two MSSM-like models: One of these models is misclassified during training, see 
figure~\ref{fig:tree_dyn}, the other one during validation. It turns out that these two MSSM-like 
models are the special $\Delta(54)$ models, where $W_3=W_4=(0^{16})$. Due to the fact that these 
models lie in a very specific area within the \Z6-II orbifold landscape, we decided to accept a 
misclassification of these models but with the benefit of obtaining a new contrast pattern that 
yields a further, significant reduction of the \Z6-II orbifold landscape. Doing so, we identify a 
new contrast pattern
\begin{equation}
c_{\text{U-sector}} ~=~ \Big\{N_{\mathrm{U}_2}^{(1)}(M) ~\geq~ 2 \;, \; N_{\mathrm{U}_3}^{(2)}(M) ~\leq~ 5 \Big\}\;,
\end{equation}
for a model $M$ to have an increased probability to be MSSM-like. We call this contrast pattern 
\emph{U-sector} as it gives bounds on the number of certain bulk matter fields, charged under 
the first or second $\E{8}$ factor, depending on $\alpha=1,2$, respectively. Using this new 
constraint on top of the previous ones, the estimated growth rate reads\footnote{Note that the 
estimated growth rate is computed based on the numbers of equivalent models. To do so, the same random 
split as for the training data of the decision tree has to be applied to the equivalent MSSM-like 
models from $D^{N_{\mathrm{sr}}}$, yielding $8\,466$ models. These numbers reduce to $8\,087$ for 
models having $G_\mathrm{SM}'$ only in the first $\E{8}$ factor and, finally, to $8\,082$ models 
fulfilling $c_{\text{U-sector}}$. Consequently, 
$\mathrm{supp}(c_{\text{U-sector}},D_{\text{MSSM-like}}^{N_{\mathrm{sr}}})=\tfrac{8082}{8087} \approx 0.999$.}
\begin{equation}\label{eq:grUsector}
\mathrm{gr}\left(c_{\text{U-sector}},D_{\text{MSSM-like}}^{N_{\mathrm{sr}}},D_{\text{\cancel{MSSM}-like}}^{N_{\mathrm{sr}}}\right) ~\approx~ \dfrac{\scriptstyle{0.999}}{\tfrac{1\,209\,687}{2\,215\,901}} ~\approx~ 1.8 ~>~ 1\;, 
\end{equation}
where $D^{N_{\mathrm{sr}}}$ is obtained by combining the datasets \emph{hidden $E_8$} and 
\emph{dynamic hidden $E_8$} from table~\ref{tab:datasets}. Some remarks on the subtleties of the 
\emph{U-sector} constraints are in order:
\begin{itemize}
\item[gr$(c)$:] Contrary to the \emph{hidden $E_8$} contrast pattern, the \emph{U-sector} 
contrast pattern can possibly exclude models which have $G_\mathrm{SM}'$ in both $\E{8}$ factors: 
in the case of $N_{\mathrm{sr}}^{(2)}(M) \geq 6$ models with $G_\mathrm{SM}'$ in both $\E{8}$ 
factors fulfill the even stronger condition $N_{\mathrm{sr}}^{(\alpha)}(M) \geq 8$ for $\alpha=1,2$. 
This can not be guaranteed for the \emph{U-sector} constraint. Therefore, the growth rate in 
eq.~\eqref{eq:grUsector} is estimated using only those models where $G_\mathrm{SM}'$ is exclusively 
in the first $\E{8}$ factor.
\item[$N_{\mathrm{U}_{2}}^{(1)}$:] Interestingly, the $\Delta(54)$ MSSM-like models excluded by the 
constraint $N_{\mathrm{U}_{3}}^{(2)}(M) \leq 5$ do obey the subsequent constraint 
$N_{\mathrm{U}_{2}}^{(1)}(M) \geq 2$ on the number of bulk matter from the $U_2$ sector and charged 
under the first $\E{8}$. Even though the decision tree decided strictly correct (by taking the 
statistics into account for optimization) and misclassified these two MSSM-like models, it is still 
appealing that all MSSM-like models (with $G_\mathrm{SM}' \in \E{8}^{(1)}$) obey this constraint. 
This observation might be worth further investigations.
\end{itemize}

\begin{figure}[t!]
\includegraphics[width=12cm]{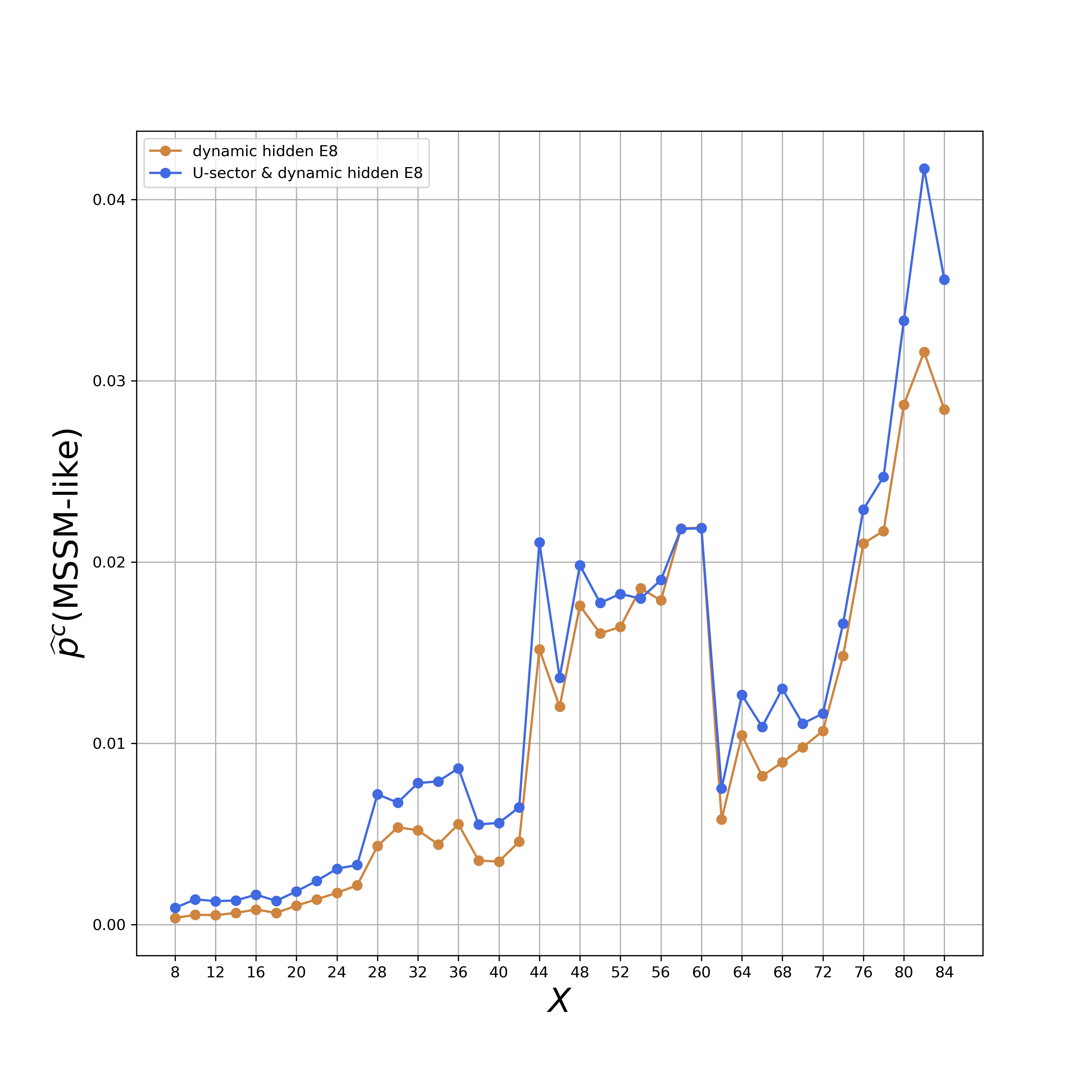}
\centering
\vspace{-1cm}
\caption{Comparison of two different \emph{dynamic hidden $E_8$} searches: (i) without and (ii) 
with the additional \emph{U-sector} constraint. For both cases, we display the probability 
$\hat{p}^{c}(\text{MSSM-like})$ (estimated on the sample) to find MSSM-like models under the 
constraint of the respective contrast pattern $c$ as a function of the threshold value $X$, where 
$c$ is (i) $N_{\mathrm{sr}}^{(2)} \geq X$ and (ii) a combination of case (i) and the 
\emph{U-sector} constraint, respectively.}
\label{fig:pmf_UvsDyn}
\end{figure}

Implementing the \emph{U-sector} contrast pattern into our search algorithm displayed in 
figure~\ref{fig:successive_create} and performing an intensive search (using all Wilson lines or 
turning off $W_5$ by hand), we obtain our final dataset, called \emph{U-sector}, see 
table~\ref{tab:datasets}. The results show once more the strength of the contrast data mining 
technique applied to the heterotic orbifold landscape: The probability to find MSSM-like models 
has increased further as shown in figure~\ref{fig:pmf_UvsDyn} and the \emph{U-sector} contrast 
pattern generalizes to the $\Z{6}$-II landscape such that we obtained many new inequivalent 
MSSM-like models. Starting from 395 inequivalent MSSM-like models in the \emph{dynamic hidden 
$E_8$} dataset we obtain now 459 models. Finally, combining all datasets yields in total 468 
inequivalent MSSM-like $\Z{6}$-II models.~\footnote{Combining these 468 MSSM-like models with 
the known models from the literature yields 481 inequivalent MSSM-like models, see 
table~\ref{tab:inequiv_ZN}.}

To summarize, we were able to significantly exceed all previous searches for MSSM-like $\Z{6}$-II 
models~\cite{Lebedev:2006kn,Lebedev:2008un,Nilles:2014owa,Olguin-Trejo:2018wpw} by excluding those 
regions in the $\Z{6}$-II orbifold landscape where most likely no MSSM-like model exists. It is 
tempting to speculate that some of our contrast patterns might even be \emph{necessary} conditions 
for \emph{all} MSSM-like models. Moreover, we learned some general features of MSSM-like models 
that can be produced in the $\Z{6}$-II orbifold landscape: We were able to identify constraints on 
physical quantities that can be interpreted and analyzed directly. Later, in 
section~\ref{sec:geometry_dependent}, we will show that the \emph{hidden $E_8$} contrast pattern 
can be transferred to other orbifold geometries while the lower bound of this constraint will be 
sensitive to the orbifold geometry under consideration.

\section[Z6-II orbifold models with Delta(54) flavor symmetry]{\boldmath $\Z{6}$-II orbifold models with $\Delta(54)$ flavor symmetry\unboldmath}
\label{sec:delta54}

Out of the 468 inequivalent MSSM-like models based on the $\Z{6}$-II orbifold geometry, there are 
two models with vanishing Wilson lines in the $\Z{3}$ torus, i.e.\ $W_3 = W_4 = (0^{16})$. Hence, 
these MSSM-like models are equipped with a $\Delta(54)$ ($R$-) flavor 
symmetry~\cite{Kobayashi:2006wq,Nilles:2012cy,Baur:2019iai}, where localized matter fields 
transform in three-dimensional representations of $\Delta(54)$.

Both models are very similar. They are based on the same shift (shift No.\ 18 according to the 
enumeration of ref.~\cite{Katsuki:1989cs}), but in different representations. This shift breaks the 
ten-dimensional $\E{8}\times\E{8}$ gauge group to 
\begin{equation}
\SO{10} \times \SU{3} \times \U{1} \quad\mathrm{and}\quad \SO{12} \times \SU{2} \times \U{1}\;,
\end{equation}
in the first and second $\E{8}$, respectively. As a remark, this shift is different from the two 
local $\SO{10}$ shifts of ref.~\cite{Lebedev:2006kn}. In a next step, $\SO{10}$ (and the hidden 
$\SO{12} \times \SU{2}$ gauge group) is broken by the Wilson lines $W_5$ and $W_6$ to the Standard 
Model gauge group, while the $\SU{3}$ factor remains unbroken as additional gauged 
$\SU{3}_\mathrm{flavor}$ flavor symmetry (Hence, the full flavor symmetry is actually 
$\SU{3}_\mathrm{flavor} \times \Delta(54)$). Consequently, both models share the same four-dimensional 
observable gauge group originating from the first $\E{8}$ factor, i.e.
\begin{equation}
\SU{3}_\mathrm{flavor} \times \SU{3}_\mathrm{C} \times \SU{2}_\mathrm{L} \times \U{1}_Y \times \U{1}^2 \times\mathrm{hidden\ sector}\;.
\end{equation}
Some details of these special MSSM-like $\Z{6}$-II orbifold models are given in the following.

\subsection[Delta(54) MSSM 1 from the Z6-II orbifold]{\boldmath $\Delta(54)$ MSSM \#1 from the $\Z{6}$-II orbifold\unboldmath}

The first $\Z{6}$-II model with $\Delta(54)$ flavor symmetry is defined by the shift vector
\begin{equation}
  V = \left(\tfrac{385}{12}, \tfrac{103}{12}, \tfrac{89}{12}, \tfrac{55}{12}, \tfrac{15}{4}, \tfrac{41}{12}, \tfrac{13}{4}, \tfrac{7}{12}\right),  \left(\tfrac{145}{4}, \tfrac{139}{12}, \tfrac{33}{4}, \tfrac{25}{4}, \tfrac{47}{12}, \tfrac{11}{4}, \tfrac{7}{4}, -\tfrac{7}{4}\right)\;,
\end{equation}
the Wilson lines $W_3 = W_4 = (0^{16})$, and 
\begin{subequations}
\begin{eqnarray}
  W_{5} & = & \left(\tfrac{17}{2},     5,     2,    -1,     1, \tfrac{5}{2},    -1,    -2\right),  \left(    5,     1,    -1,     0,     1, \tfrac{1}{2},     0, \tfrac{3}{2}\right)\;,\\
  W_{6} & = & \left(\tfrac{1}{2}, -\tfrac{3}{2}, -\tfrac{1}{2},     3,     2, \tfrac{3}{2},     0,     1\right),  \left(    2, -\tfrac{3}{2}, \tfrac{1}{2},    -1,     2, \tfrac{7}{2},     0, -\tfrac{7}{2}\right)\;.
\end{eqnarray}
\end{subequations}
It is called \emph{MSSM431} in the model-file ``Z6-II\_1\_1.txt''~\cite{Parr:2019anc} that can be 
loaded to the \texttt{orbifolder}. For his model, the hidden gauge group is broken to
\begin{equation}
\SU{3} \times \SU{3} \times \U{1}^4\;.
\end{equation}
The massless string spectrum is summarized in table~\ref{tab:Z6IIDelta54Model431}. Interestingly, 
the spectrum contains flavons that are triplets under both $\Delta(54)$ and $\SU{3}_\mathrm{flavor}$. 
Hence, their vacuum expectation values could break the full flavor group to a diagonal subgroup.

\begin{table}[t!]
\center
\begin{tabular}{|r|l|c||r|l|c|}
\hline
 \# & irrep & labels & \# & irrep & labels \\
\hline
\hline
    6 & $\left( \rep{1}; \rep{3},\rep{2}; \rep{1}, \rep{1}\right)_{-\tfrac{1}{6}}$ & $q_{i}$ &
    3 & $\left( \rep{1};\crep{3},\rep{2}; \rep{1}, \rep{1}\right)_{ \tfrac{1}{6}}$ & $\bar{q}_{i}$\\
    1 & $\left( \rep{3};\crep{3},\rep{1}; \rep{1}, \rep{1}\right)_{ \tfrac{2}{3}}$ & $\bar{u}_{i}$ & & & \\
    1 & $\left(\crep{3};\crep{3},\rep{1}; \rep{1}, \rep{1}\right)_{-\tfrac{1}{3}}$ & $\bar{d}_{i}$ & & & \\
    5 & $\left( \rep{1};\crep{3},\rep{1}; \rep{1}, \rep{1}\right)_{-\tfrac{1}{3}}$ & $\bar{d}_{i}$ &
    5 & $\left( \rep{1}; \rep{3},\rep{1}; \rep{1}, \rep{1}\right)_{ \tfrac{1}{3}}$ & $d_{i}$ \\
   10 & $\left( \rep{1}; \rep{1},\rep{2}; \rep{1}, \rep{1}\right)_{ \tfrac{1}{2}}$ & $\ell_{i}$ &
    7 & $\left( \rep{1}; \rep{1},\rep{2}; \rep{1}, \rep{1}\right)_{-\tfrac{1}{2}}$ & $\bar{\ell}_{i}$\\   
    1 & $\left( \rep{3}; \rep{1},\rep{1}; \rep{1}, \rep{1}\right)_{           -1}$ & $\bar{e}_{1}
    $ & & & \\
\hline
   12 & $\left( \rep{1}; \rep{1},\rep{1}; \rep{1}, \rep{1}\right)_{ \tfrac{1}{2}}$ & $s_{i}^{+}$ &
   12 & $\left( \rep{1}; \rep{1},\rep{1}; \rep{1}, \rep{1}\right)_{-\tfrac{1}{2}}$ & $s_{i}^{-}$\\
    4 & $\left( \rep{1}; \rep{1},\rep{1}; \rep{1},\crep{3}\right)_{ \tfrac{1}{2}}$ & $s_{i}^{+}$ &
    2 & $\left( \rep{1}; \rep{1},\rep{1}; \rep{1}, \rep{3}\right)_{-\tfrac{1}{2}}$ & $s_{i}^{-}$\\
    2 & $\left( \rep{1}; \rep{1},\rep{1}; \rep{1}, \rep{3}\right)_{ \tfrac{1}{2}}$ & $s_{i}^{+}$ &
    4 & $\left( \rep{1}; \rep{1},\rep{1}; \rep{1},\crep{3}\right)_{-\tfrac{1}{2}}$ & $s_{i}^{-}$\\
    4 & $\left( \rep{1}; \rep{1},\rep{1};\crep{3}, \rep{1}\right)_{ \tfrac{1}{2}}$ & $s_{i}^{+}$ &
    2 & $\left( \rep{1}; \rep{1},\rep{1}; \rep{3}, \rep{1}\right)_{-\tfrac{1}{2}}$ & $s_{i}^{-}$\\
    2 & $\left( \rep{1}; \rep{1},\rep{1}; \rep{3}, \rep{1}\right)_{ \tfrac{1}{2}}$ & $s_{i}^{+}$ &
    4 & $\left( \rep{1}; \rep{1},\rep{1};\crep{3}, \rep{1}\right)_{-\tfrac{1}{2}}$ & $s_{i}^{-}$\\
\hline
   11 & $\left(\crep{3}; \rep{1},\rep{1}; \rep{1}, \rep{1}\right)_{            0}$ & $f_{i}$ &
   10 & $\left( \rep{3}; \rep{1},\rep{1}; \rep{1}, \rep{1}\right)_{            0}$ & $\bar{f}_{i}$\\   
   17 & $\left( \rep{1}; \rep{1},\rep{1}; \rep{1}, \rep{1}\right)_{            0}$ & $s_{i}^{0}$ & & & \\
    7 & $\left( \rep{1}; \rep{1},\rep{1}; \rep{3}, \rep{1}\right)_{            0}$ & $s_{i}^{0}$ &
    6 & $\left( \rep{1}; \rep{1},\rep{1};\crep{3}, \rep{1}\right)_{            0}$ & $s_{i}^{0}$\\
    7 & $\left( \rep{1}; \rep{1},\rep{1}; \rep{1}, \rep{3}\right)_{            0}$ & $s_{i}^{0}$ & 
    6 & $\left( \rep{1}; \rep{1},\rep{1}; \rep{1},\crep{3}\right)_{            0}$ & $s_{i}^{0}$\\
    1 & $\left( \rep{1}; \rep{1},\rep{1};\crep{3}, \rep{3}\right)_{            0}$ & $s_{i}^{0}$ &
    1 & $\left( \rep{1}; \rep{1},\rep{1}; \rep{3},\crep{3}\right)_{            0}$ & $s_{i}^{0}$\\
    1 & $\left( \rep{1}; \rep{1},\rep{1}; \rep{3}, \rep{3}\right)_{            0}$ & $s_{i}^{0}$ & & &  \\
\hline
\end{tabular}
\caption{Massless matter spectrum of the first $\Z{6}$-II model with $\Delta(54)$ flavor symmetry. 
Many SM matter fields build three-dimensional representations of $\Delta(54)$: For instance, all 6 
quark-doublets $q_i$ combine to two three-dimensional representations of $\Delta(54)$, similar for 
all partners $\bar{q}_{i}$, 6 out of 10 lepton-doublets (or down-Higgs) $\ell_{i}$, and 3 out of 7 
anti-lepton-doublets (or up-Higgs) $\bar{\ell}_{i}$. Furthermore, several SM singlets $s_{i}^{0}$ 
are three-dimensional representations of $\Delta(54)$. Finally, the SM singlets $f_{i}$ and 
$\bar{f}_{i}$ are flavons of $\SU{3}_\mathrm{flavor}$ (interestingly, some of these flavons are 
simultaneously triplets of $\Delta(54)$).}
\label{tab:Z6IIDelta54Model431}
\end{table}

\subsection[Delta(54) MSSM 2 from the Z6-II orbifold]{\boldmath $\Delta(54)$ MSSM \#2 from the $\Z{6}$-II orbifold\unboldmath}

The second $\Z{6}$-II model with $\Delta(54)$ flavor symmetry is given by the shift vector
\begin{equation}
V = \left( \tfrac{247}{4}, \tfrac{69}{4}, \tfrac{61}{4}, \tfrac{137}{12}, \tfrac{101}{12}, \tfrac{65}{12}, \tfrac{11}{4}, -\tfrac{1}{4} \right)  \left( \tfrac{167}{3}, \tfrac{83}{6}, \tfrac{34}{3}, \tfrac{55}{6}, \tfrac{47}{6}, \tfrac{29}{6}, \tfrac{13}{6}, \tfrac{1}{6} \right)\;,
\end{equation}
the Wilson lines $W_3 = W_4 = (0^{16})$, and 
\begin{subequations}
\begin{eqnarray}
W_5 & = & \left(34, \tfrac{19}{2}, \tfrac{15}{2}, \tfrac{11}{2}, \tfrac{11}{2}, \tfrac{7}{2}, \tfrac{1}{2}, -3 \right)
 \left( -\tfrac{113}{4}, -\tfrac{23}{4}, -\tfrac{17}{4}, -\tfrac{13}{4}, -\tfrac{27}{4}, -\tfrac{17}{4}, -\tfrac{3}{4}, \tfrac{5}{4} \right)\;,\\
W_6 & = & \left(-\tfrac{11}{4}, -\tfrac{1}{4}, \tfrac{5}{4}, -\tfrac{7}{4}, -\tfrac{11}{4}, \tfrac{1}{4}, -\tfrac{1}{4}, -\tfrac{11}{4}\right)
 \left(   16,     1,     3,     2,     4, \tfrac{5}{2}, -\tfrac{1}{2},    -2\right)\;.
\end{eqnarray}
\end{subequations}
It is called \emph{MSSM438} in the model-file ``Z6-II\_1\_1.txt''~\cite{Parr:2019anc}. In this 
case, the hidden gauge group is broken to
\begin{equation}
\SU{5} \times \U{1}^4\;,
\end{equation}
in four dimensions and the massless string spectrum is given in 
table~\ref{tab:Z6IIDelta54Model438} in appendix~\ref{app:table}. Compared to the $\Delta(54)$ MSSM 
\#1, $\SU{3}_\mathrm{flavor}$ and $\Delta(54)$ seem to have interchanged their role for many 
representations of quarks and leptons.

\section{Geometry-dependent contrast patterns}
\label{sec:geometry_dependent}

The results from the previous sections were developed in the $\Z{6}$-II $(1, 1)$ orbifold geometry. 
However, the basic insights from this analysis can be transferred easily to other orbifold 
geometries. Foremost, the concept of a \emph{hidden $E_8$} contrast pattern can be applied directly 
to other orbifold geometries: The number of unbroken roots from the hidden $\E{8}$ is computed 
identically for all orbifold geometries and does not depend on some unknown sorting. Unfortunately, 
this is not given for the \emph{U-sector} contrast pattern: The number of bulk matter fields 
$N_{\mathrm{U}_{a}}^{(\alpha)}$ for $a=1,2,3$ depends on the twist vectors of a given orbifold 
geometry, see eq.~\eqref{eq:number_U_sector}. Then, the sorting of $N_{\mathrm{U}_{1}}^{(\alpha)}$, 
$N_{\mathrm{U}_{2}}^{(\alpha)}$ and $N_{\mathrm{U}_{3}}^{(\alpha)}$ is determined by the sorting of 
the entries in the twist vectors, which is typically sorted from from small to large rotation 
angles. However, it is not clear why a particular U-sector might be special among the different 
sectors, i.e.\ if a special status of an U-sector is related to the sorting or to some nontrivial 
relation between all sectors.

Therefore, we begin with the \emph{dynamic hidden $E_8$} search and analyze the U-sector later. In 
order to apply the \emph{dynamic hidden $E_8$} contrast pattern to all \Z{N} orbifold geometries, 
we first have to identify the lower bound of $N_{\mathrm{sr}}^{(2)}(M)$ (being 6 in 
eq.~\eqref{eq:FirstCP}) for each $\Z{N}$ orbifold geometry $\mathbbm{O}$. Thus, we define
\begin{align}
X_{\mathrm{min}}(\mathbbm{O}) ~=~
\mathrm{min}\left( \Big\{N_{\mathrm{sr}}^{(2)}(M)
 \ \Big| \ M \in D_{(\mathbbm{O}\text{ from \cite{Nilles:2014owa,Olguin-Trejo:2018wpw}})} \Big\}
 \right)\;.
\end{align}
To compute these bounds, we use the traditional searches of refs.~\cite{Nilles:2014owa,Olguin-Trejo:2018wpw} 
as a background search and split the combined dataset into datasets 
$D_{(\mathbbm{O}\text{ from \cite{Nilles:2014owa,Olguin-Trejo:2018wpw}})}$ corresponding to the 
different \Z{N} orbifold geometries $\mathbbm{O}$. Then, the conventional search in 
refs.~\cite{Nilles:2014owa,Olguin-Trejo:2018wpw} can be seen as a search with
$N_{\mathrm{sr}}^{(2)} \geq 0$ in our approach. Thus, we can analyze the results of the traditional 
search~\cite{Nilles:2014owa,Olguin-Trejo:2018wpw} to obtain the lower bounds 
$X_{\mathrm{min}}(\mathbbm{O})$ for all $\Z{N}$ orbifold geometries. The 
results are stated in table~\ref{tab:ZN_hiddenE8}. Moreover, the background datasets 
$D_{(\mathbbm{O}\text{ from \cite{Nilles:2014owa,Olguin-Trejo:2018wpw}})}$ allow us to focus the 
\emph{dynamic hidden $E_8$} contrast pattern on thresholds greater than 
$X_{\mathrm{min}}(\mathbbm{O})$, i.e.\ $N_{\mathrm{sr}}^{(2)} > X_{\mathrm{min}}(\mathbbm{O})$, in 
order to save computational resources in our search.
\begin{table}[t!]
\center
\renewcommand{\arraystretch}{1.5}
\begin{tabular}{|c|c|c|c|c|ccc|cc|c|c|}
\hline
orbifold & \Z4 &\Z6-I & \Z6-II  &\Z7 & \Z8-I     &          &          &\Z8-II    &          &\Z{12}-I        &\Z{12}-II  \\
\cline{2-12}
geometry $\mathbbm{O}$ & all & all & all & (1,1) & (1,1) & (2,1)& (3,1) & (1,1) & (2,1) & all & (1,1)  \\
 \hline
$X_{\mathrm{min}}(\mathbbm{O})$ & 4  & 12 & 6 & 56 & 6  & 6  & 4  & 0  & 4  &  4 &  6  \\
 \hline
\end{tabular}
\caption{Lower bounds $X_{\mathrm{min}}(\mathbbm{O})$ on the number $N_{\mathrm{sr}}^{(2)}$ of 
unbroken roots from the hidden $\E{8}$ factor for MSSM-like orbifold models from 
refs.~\cite{Nilles:2014owa,Olguin-Trejo:2018wpw} for various \Z{N} orbifold geometries $\mathbbm{O}$ 
(where ``all'' refers to the various lattices for a given $\Z{N}$ orbifold geometry, as given in 
the first column of table~\ref{tab:inequiv_ZN}.).
}
\label{tab:ZN_hiddenE8}
\end{table} 

First, let us state the results of our search in table~\ref{tab:inequiv_ZN}: For each $\Z{N}$ 
orbifold geometry, we give the numbers of inequivalent MSSM-like orbifold models that we found 
using our \emph{dynamic hidden $E_8$} contrast pattern and compare these numbers to the 
literature. Several remarks are in order. 
\begin{table}[t]
\center
\renewcommand{\arraystretch}{1.0}
\begin{tabular}{|cc||r|r|r||r|}
\multicolumn{6}{c}{\textbf{inequivalent MSSM-like orbifold models}} \\
\hline
\multicolumn{2}{|c||}{orbifold} & \multicolumn{1}{c|}{\# MSSM-like}               & \multicolumn{1}{c|}{\# MSSM-like}                     & \multicolumn{1}{c||}{\# MSSM-like using} & \multicolumn{1}{c|}{\# MSSM-like}\\ 
\multicolumn{2}{|c||}{geometry} & \multicolumn{1}{c|}{from~\cite{Nilles:2014owa}} & \multicolumn{1}{c|}{from~\cite{Olguin-Trejo:2018wpw}} & \multicolumn{1}{c||}{contrast patterns}  & \multicolumn{1}{c|}{`merged'}\\
\hline
\hline
\Z4       & (2,1) & 128 &    138 &    125 &    179 \\
          & (3,1) &  25 &     26 &     33 &     33 \\
\hline                
\Z6-I     & (1,1) &  31 &     30 &     31 &     31 \\
          & (2,1) &  31 &     30 &     31 &     31 \\
\hline
\Z6-II    & (1,1) & 348 &    363 &    468 &    481 \\
          & (2,1) & 338 &    349 &    395 &    443 \\
          & (3,1) & 350 &    351 &    415 &    482 \\
          & (4,1) & 334 &    354 &    407 &    464 \\ 
\hline
\Z7       & (1,1) &   0 &      1 &      1 &      1 \\
\hline
\Z8-I     & (1,1) & 263 &    256 &    248 &    271 \\
          & (2,1) & 164 &    155 &    144 &    164 \\
          & (3,1) & 387 &    377 &    408 &    430 \\ 
\hline
\Z8-II    & (1,1) & 638 & 1\,833 & 1\,259 & 2\,289 \\
          & (2,1) & 260 &    489 &    349 &    555 \\  
\hline
\Z{12}-I  & (1,1) & 365 &    556 &    610 &    625 \\
          & (2,1) & 385 &    554 &    607 &    625 \\ 
\hline
\Z{12}-II & (1,1) & 211 &    352 &    365 &    435 \\
\hline  
\end{tabular}
\caption{Table of inequivalent MSSM-like orbifold models for all $\Z{N}$ orbifold geometries, see 
also~\cite{Parr:2019anc}. Note that the numbers of MSSM-like orbifold models listed in the third 
column differ from those in ref.~\cite{Olguin-Trejo:2018wpw}. This is due to an improvement of the 
\texttt{orbifolder} which has led to a better comparison of models and identified some duplicates 
in these sets. The last column, gives our final results: the numbers of inequivalent MSSM-like 
orbifold models obtained by merging the three datasets of the previous columns.}
\label{tab:inequiv_ZN}
\end{table} 
One can observe that the \emph{dynamic hidden $E_8$} search was able to find many new inequivalent 
MSSM-like orbifold models in almost all orbifold geometries. Foremost, the different \Z6-II 
orbifold geometries as well as the \Z{12}-I case have improved strongly using our contrast 
patterns. Note that for \Z6-II $(1, 1)$, the 13 additional MSSM-like models 
from refs.~\cite{Nilles:2014owa,Olguin-Trejo:2018wpw} fulfill all our constraints derived in 
section~\ref{sec:contrast_pattern}. Consequently, even though these models were missed in our 
search, they are part of our search area. Hence, these models would have been found in an extended 
search. A great success of our contrast patterns is also given by the appearance of the MSSM-like 
\Z7 model. This model was found so far only in refs.~\cite{RamosSanchez:2008tn,Olguin-Trejo:2018wpw} using an 
orbifold-specific search strategy, as described in appendix A of ref.~\cite{RamosSanchez:2008tn}. 
Also the \Z6-I orbifold geometry is remarkable: In this case, we find a huge amount of equivalent 
MSSM-like models but only 31 inequivalent ones remain. These 31 inequivalent models were found very 
easily by searching in areas of the \Z6-I orbifold landscape with large hidden sector gauge groups, 
c.f.\ the lower bound $X_{\mathrm{min}}(\Z6\text{-I})=12$ given in table~\ref{tab:ZN_hiddenE8}. 
Note that the lower bounds are computed for those models where $G_\mathrm{SM}'$ appears in one 
$\E{8}$ factor only. While most of the bounds in table~\ref{tab:ZN_hiddenE8} are weaker than 
$N_{\mathrm{sr}}^{(1)}\geq 8$ from section~\ref{sec:constraint_G_sm}, we have to be careful in the 
cases of both \Z6-I orbifold geometries. There, we find a lower bound 
$X_{\mathrm{min}}($\Z6\text{-I}$) = 12$. Thus, our search algorithm could in principle miss 
MSSM-like $\Z6$-I models where each $\E{8}$ factor contains $G_\mathrm{SM}'$. We analyze these 
problematic models separately and find a lower bound $X_{\mathrm{min}}($\Z6-I$)=10$ for these cases 
(because $(N_{\mathrm{sr}}^{(1)},N_{\mathrm{sr}}^{(2)})$ takes only the values $(8,12)$ and 
$(10,10)$ in the background dataset). This means that MSSM-like $\Z6$-I models which have 
$G_\mathrm{SM}'$ in both $\E{8}$ factors are contained in our search $N_{\mathrm{sr}}^{(2)} \geq 10$ 
for both \Z6-I orbifold geometries.

Furthermore, it seems that for some orbifold geometries like \Z{8}-II $(1, 1)$ the conventional 
approach has some advantages. However, a comparison is difficult since it is not known how much 
computational power was invested to obtain these numbers. Moreover, for \Z{8}-II $(1, 1)$ our 
contrast patterns seem to be less efficient since there is no lower bound for $N_{\mathrm{sr}}^{(2)}$, 
i.e.\ MSSM-like models with $N_{\mathrm{sr}}^{(2)}=0$ exist for \Z{8}-II $(1, 1)$, and there are 
many inequivalent MSSM-like models for low values of $N_{\mathrm{sr}}^{(2)}$, which can be reached 
by the conventional search algorithm as well. Hence, on first sight it seems that our search algorithm 
is too complex for such geometries and the additional effort in computing constraints is not 
rewarded. However, this conclusion is premature: The merged datasets in table~\ref{tab:inequiv_ZN} 
show that our contrast patterns could still significantly improve the numbers of inequivalent 
MSSM-like orbifold models in these geometries. Thus, our search algorithm was able to find new 
MSSM-like orbifold models in corners of the landscape that were missed by the conventional 
approach.

On the basis of the `merged' datasets we can now use decision trees to derive the \emph{U-sector} 
constraints as in the case of the $\Z{6}$-II orbifold geometry. The results are given in 
table~\ref{tab:ZN_usector}.
\begin{table}[t]
\center
\renewcommand{\arraystretch}{1.1}
\begin{tabular}{|cr|rrr|rrr|r|}
\hline
\multicolumn{2}{|c|}{orbifold}  &  
\multirow{2}{*}{$N_{\mathrm{U}_{1}}^{(1)}$} & \multirow{2}{*}{$N_{\mathrm{U}_{2}}^{(1)}$} & \multirow{2}{*}{$N_{\mathrm{U}_{3}}^{(1)}$} & \multirow{2}{*}{$N_{\mathrm{U}_{1}}^{(2)}$} & \multirow{2}{*}{$N_{\mathrm{U}_{2}}^{(2)}$} & \multirow{2}{*}{$N_{\mathrm{U}_{3}}^{(2)}$}  & gr(c) 
\\
\multicolumn{2}{|c|}{geometry}  & & & & & &  &
   \\
\hline
\hline
\Z4    & (2,1) &   & & $ \geq 4 $ & & & $\leq 1$      & 5.32   \\ 
       & (3,1) &   & & $ \geq 4 $ & & & $\leq 1$      & 6.92   \\ 
\hline                
\Z6-I  & (1,1) &  $\geq 13$ &   & $\geq 14$ &  & &    & 2.79   \\ 
       & (2,1) &  $\geq 13$ &   & $\geq 14$ &  & &    & 2.78    \\
\hline
\Z6-II & (1,1) &    & $\geq 2$ & & & & $\leq 5$       & 1.81   \\ 
       & (2,1) &    & $\geq 2$ & & & & $\leq 5$       & 1.60   \\ 
       & (3,1) &    & $\geq 2$ & & & & $\leq 5$       & 1.70   \\ 
       & (4,1) &    & $\geq 2$ & & & & $\leq 5$       & 1.86   \\  
\hline
\Z8-I  & (1,1) &  & & $ \geq 4 $ & & & $\leq 25 $     & 1.22   \\ 
       & (2,1) &    & & $ \geq 4 $ & & & $\leq 25$    & 1.23   \\
       & (3,1) &    & & $ \geq 8 $ & & &              & 2.21   \\
\hline
\Z8-II & (1,1) &    & & $ \geq 4 $ & & & $\leq 41$    & 1.61   \\
       & (2,1) &    & & $\leq 3 $ & & & $\leq 1 $     & 1.78   \\
       &       &    & & $ \geq 4  $ & & & $\leq 41 $  & 1.01   \\
\hline
\Z{12}-I  & (1,1) &  $\leq 10 $ & & $ \geq 2 $ & & &     &  1.24  \\ 
          & (2,1) &  $\leq 10 $ & & $ \geq 2 $ & & &     &  1.24  \\ 
\hline
\Z{12}-II & (1,1) &  $\geq 2$ & & & & & $\leq 5$      & 1.71   \\
\hline  
\end{tabular}
\caption{U-sector constraints for various $\Z{N}$ orbifold geometries, based on the merged 
datasets of table~\ref{tab:inequiv_ZN}. We neglect the $\Z{7}$ orbifold geometry because there is 
only one MSSM-like model available.}
\label{tab:ZN_usector}
\end{table} 
Let us analyze the resulting \emph{U-sector} contrast patterns in some detail: In nearly all 
orbifold geometries it is possible to get a recall in the validation set of 1.00 and no MSSM-like 
model is missed in the training set. Only for the \Z6-II orbifold geometries $(1, 1)$, $(2, 1)$ and 
$(3, 1)$ a very few \emph{false negative} predictions were made either in the validation set or in 
the training set. Another special case is the \Z8-II $(2, 1)$ orbifold geometry: In order to get a 
growth rate larger than one the decision tree had to split the set of MSSM-like models into two 
sets at the first node with a constraint on $N_{\mathrm{U}_{3}}^{(1)}$. Then, for both sets a 
second split takes $N_{\mathrm{U}_{3}}^{(2)}$ into account, resulting in a growth rate of 1.78 for 
the first set (containing only 3\% of the MSSM-like \Z8-II $(2, 1)$ models from the training set) 
and a growth rate of 1.01 for the second set (containing 97\% of the models), respectively. In 
this context, let us mention that for \Z8-I $(1, 1)$ it is possible to create another decision 
tree with a constraint ${c'}_{\mathrm{U-sector}} = \big\{ N_{\mathrm{U}_{3}}^{(1)} \geq 8, N_{\mathrm{U}_{3}}^{(2)} \leq 25 \big\}$ 
that has a recall value of 1.00 and gr$({c'}_{\mathrm{U-sector}})= 2.18$, however, with the 
trade-off of missing one MSSM-like model from the training data.
 
Interestingly, it seems that the \emph{U-sector} constraints in table~\ref{tab:ZN_usector} show 
some patterns on their own. Foremost, it is remarkable that for a given twist vector the exact 
orbifold geometry (i.e.\ the choice of the six-torus which is enumerated by the number $l=1,2,\ldots$ 
in the label $(l,1)$) does not have any significant effect. This could be used to extrapolate from one 
orbifold geometry to another. If the constraints are different within one twist vector one should 
be careful and probably take the weakest constraint, e.g.\ $N_{\mathrm{U}_{1}}^{(3)} \geq 4$ for 
\Z8-I. This can be seen as regularizing the machine learning model. One can use this insight to 
avoid overfitting and to use more statistics from other orbifold geometries. Additionally, one 
can observe that the hidden sector is completely dominated by a `$\leq$' constraint in the 
\emph{U$_{3}$-sectors}, while the visible sector favors `$\geq$' (except for \Z{12}-I and \Z6-I). 
This shows that there is still more structure to explore in the heterotic orbifold landscape and, 
more importantly, that we are on the right track to obtain necessary conditions on both the 
observable $\E{8}$ factor, containing the MSSM, and the hidden $\E{8}$ factor.

\section{Conclusion}
\label{sec:conclusion}

In this paper, we have developed an advanced search strategy for MSSM-like orbifold models using 
the $\Z{6}$-II $(1, 1)$ orbifold geometry as a test case. We obtained a significant improvement 
from 363 inequivalent MSSM-like models~\cite{Olguin-Trejo:2018wpw} to 481, see 
table~\ref{tab:inequiv_ZN}. To do so, we used a technique called contrast data mining, where one 
identifies so-called contrast patterns that help to distinguish between MSSM-like models and others. 
In principle, this technique is easy to generalize to all orbifold geometries and, presumably, to 
other string compactifications. As a first step towards this, we analyzed all $\Z{N}$ orbifold 
geometries in section~\ref{sec:geometry_dependent} and showed that in all cases our contrast 
patterns significantly enhance the known datasets of MSSM-like orbifold models, see 
table~\ref{tab:inequiv_ZN}. Let us stress that this new search strategy is superior by orders of 
magnitudes with respect to the computing time. Theoretically, the conventional search algorithm can 
find \emph{all} MSSM-like orbifold models. However, this would correspond to an unfeasible amount 
of computing time because the effort for finding a new MSSM-like model grows exponentially with the 
number of already constructed models. This fact was studied in detail in ref.~\cite{Dienes:2006ca} 
and can also be inferred from figures~\ref{fig:hist_spec_MI},~\ref{fig:inequ_Repetition_vs_SurvRoots} 
and~\ref{fig:inequ_Repetition_vs_SurvRoots_dyn}. These figures show that the towers of already 
known orbifold models dominate the search and statistically keep growing before new orbifold models 
are expected to appear. Hence, with increasing search time the probability to find a new orbifold 
model is suppressed further and further. 

Consequently, we believe that contrast patterns can be of great importance when studying the 
string landscape. In addition, contrast patterns are particularly useful as they have a clear physical 
interpretation. In our setup of heterotic orbifolds, we identified the following contrast patterns: 
the number of unbroken roots in the hidden $\E{8}$ factor and the numbers of various bulk matter 
fields, charged under first or second $\E{8}$ factor. Hence, our contrast patterns are related to 
bulk fields that originate from the compactification of the ten-dimensional $\E{8}\times\E{8}$ 
gauge bosons. Moreover, our contrast patterns have direct phenomenological implications, as they 
are important for supersymmetry breaking via hidden sector gaugino 
condensation~\cite{Dienes:2006ut,Lebedev:2006tr}, gauge-Higgs 
unification~\cite{Hall:2001zb,Kubo:2001zc} and gauge-top unification~\cite{Hosteins:2009xk}. 
Further studies along these lines have to follow.

Moreover, using the approach with contrast patterns it was possible to solve some long standing 
issues in the heterotic orbifold landscape, namely:
\begin{itemize}
\item We found many new MSSM-like orbifold models, especially in corners of the heterotic orbifold 
landscape that were hardly accessible by the conventional search algorithms, see 
table~\ref{tab:inequiv_ZN}.
\item As stated in table~\ref{tab:datasets}, it was possible to proof the existence of MSSM-like 
$\Z{6}$-II models with vanishing Wilson line of order three, i.e.\ $W_{3}=W_{4}=(0^{16})$. This is 
the first time that such models are described in the literature. They might be phenomenologically 
interesting as they are equipped with a $\Delta(54)$ flavor symmetry, see section~\ref{sec:delta54}.
\item Furthermore, using the new technique we were able to reproduce the only known MSSM-like model 
in the $\Z{7}$ orbifold geometry~\cite{RamosSanchez:2008tn,Olguin-Trejo:2018wpw}. This model could 
not be found by any random search so far. Instead, it was found using a method (described in 
appendix A of ref.~\cite{RamosSanchez:2008tn}) that is not feasible for most other orbifold 
geometries.
\item Moreover, even though the main aim of our search algorithm is to find \emph{inequivalent} 
MSSM-like orbifold models, we obtain in addition an important byproduct: our contrast patterns 
significantly increase the probability to find MSSM-like orbifold models in certain regions of the 
heterotic orbifold landscape. In future applications, the models that are classified as 
inequivalent by the \texttt{orbifolder} may differ in some other aspects, e.g.\ in their Yukawa 
couplings, see section~\ref{sec:orbifolder}. As soon as a preferred MSSM-like orbifold model is 
identified, our search algorithm allows to explore a specific part of the landscape in order to 
find models that have similar spectra but are not necessarily equivalent with respect to the full model.
\item Also this work can be seen to be a fundamental step in applying further machine learning 
techniques to the heterotic orbifold landscape. In this paper, we are fighting the imbalance of the 
string theory datasets, i.e.\ we are trying to get enough data of the minority class, which is build up by 
the MSSM-like models. Especially, deep learning techniques are very sensible to unbalanced data and 
tend to perform worse than traditional machine learning methods.
\end{itemize}

Finally, we want to state some preferred properties of contrast pattern that should be kept in mind 
when constructing new features in the future. These properties are useful for implementation as 
well as for the impact of a new contrast pattern. The most important property is that a new feature 
can be checked quickly and easily, since it will be computed several times during the successive 
search algorithm. In addition, it is an advantage if a contrast pattern is testable at each step of 
the successive construction, see figure~\ref{fig:successive_create}. Therefore, a new feature must 
be a monotonically decreasing (increasing) function with respect to the successive creation of shifts 
and Wilson lines. Moreover, in combination with the monotonic behavior the constraint has to be a 
lower (upper) bound on the model. For example, the minimal number of unbroken roots is a good 
contrast pattern since it can only decrease at each step in figure~\ref{fig:successive_create} and, 
therefore, it is a monotonically decreasing function with a lower bound. On the other side, the 
maximal number of bulk fields in a certain \emph{$U$-sector} can only be checked at the last step 
in figure~\ref{fig:successive_create} since this number is also a monotonically decreasing function 
and any subsequently chosen Wilson line can decrease this value further. Hence, even though the 
number of bulk fields is a monotonically decreasing function with respect to the successive 
creation of shifts and Wilson lines, the \emph{$U$-sector} contrast pattern is given by an upper 
bound, which weakens this contrast pattern.

In conclusion, this paper shows that techniques from data mining and machine learning can be 
applied successfully to the heterotic orbifold landscape and produce practical results, i.e.\ 
novel MSSM-like models that were out of reach using all traditional approaches so far. Further 
investigations in this direction have to be done in order to complete the set of contrast 
patterns. One might speculate that contrast patterns could ultimately help to identify an analytic 
formula for the construction of MSSM-like models in the heterotic orbifold landscape.

\section{Acknowledgements}
This work was supported by the Deutsche Forschungsgemeinschaft (SFB1258). We would like to thank Sa{\'u}l Ramos-S{\'a}nchez for useful discussions.

\appendix

\section[Spectrum of Z6-II MSSM 2 with Delta(54) flavor symmetry]{\boldmath Spectrum of $\Z{6}$-II MSSM \# 2 with $\Delta(54)$ flavor symmetry\unboldmath}
\label{app:table}

\begin{table}[h!]
\center
\begin{tabular}{|r|l|c||r|l|c|}
\hline
 \# & irrep & labels & \# & irrep & labels \\
\hline
\hline
 1 & $\left( \rep{3};\crep{3},\rep{2};  \rep{1}\right)_{ \tfrac{1}{6}}$ & $ q_i$ & & & \\
 6 & $\left( \rep{1}; \rep{3},\rep{1};  \rep{1}\right)_{-\tfrac{2}{3}}$ & $\bar{u}_i$ & 
 3 & $\left( \rep{1};\crep{3},\rep{1};  \rep{1}\right)_{ \tfrac{2}{3}}$ & $u_i$\\   
 1 & $\left(\crep{3}; \rep{3},\rep{1};  \rep{1}\right)_{ \tfrac{1}{3}}$ & $\bar{d}_i$ & & & \\
 4 & $\left( \rep{1}; \rep{3},\rep{1};  \rep{1}\right)_{ \tfrac{1}{3}}$ & $\bar{d}_i$ &
 4 & $\left( \rep{1};\crep{3},\rep{1};  \rep{1}\right)_{-\tfrac{1}{3}}$ & $d_i$ \\
11 & $\left( \rep{1}; \rep{1},\rep{2};  \rep{1}\right)_{-\tfrac{1}{2}}$ & $\ell_{i}$ &
 8 & $\left( \rep{1}; \rep{1},\rep{2};  \rep{1}\right)_{ \tfrac{1}{2}}$ & $\bar{\ell}_{i}$\\
 6 & $\left( \rep{1}; \rep{1},\rep{1};  \rep{1}\right)_{            1}$ & $\bar{e}_{i}$ &
 3 & $\left( \rep{1}; \rep{1},\rep{1};  \rep{1}\right)_{           -1}$ & $e_{i}$ \\
\hline
 3 & $\left( \rep{1};\crep{3},\rep{1};  \rep{1}\right)_{ \tfrac{1}{6}}$ & $\bar{v}_{i}$ &
 3 & $\left( \rep{1}; \rep{3},\rep{1};  \rep{1}\right)_{-\tfrac{1}{6}}$ & $v_{i}$ \\
12 & $\left( \rep{1}; \rep{1},\rep{1};  \rep{1}\right)_{ \tfrac{1}{2}}$ & $s_{i}^{+}$ & 
 3 & $\left(\crep{3}; \rep{1},\rep{1};  \rep{1}\right)_{-\tfrac{1}{2}}$ & $s_{i}^{-}$ \\ 
 2 & $\left(\crep{3}; \rep{1},\rep{1};  \rep{1}\right)_{ \tfrac{1}{2}}$ & $s_{i}^{+}$ &
 3 & $\left( \rep{3}; \rep{1},\rep{1};  \rep{1}\right)_{-\tfrac{1}{2}}$ & $s_{i}^{i}$ \\
15 & $\left( \rep{1}; \rep{1},\rep{2};  \rep{1}\right)_{            0}$ & $m_{i}$ & & & \\
 2 & $\left( \rep{1}; \rep{1},\rep{2};  \rep{5}\right)_{            0}$ & $m_{i}$ &
 1 & $\left( \rep{1}; \rep{1},\rep{2}; \crep{5}\right)_{            0}$ & $m_{i}$ \\   
\hline
25 & $\left( \rep{1}; \rep{1},\rep{1};  \rep{1}\right)_{            0}$ & $s_{i}^{0}$  & & & \\ 
11 & $\left(\crep{3}; \rep{1},\rep{1};  \rep{1}\right)_{            0}$ & $\bar{f}_{i}$ &
10 & $\left( \rep{3}; \rep{1},\rep{1};  \rep{1}\right)_{            0}$ & $f_{i}$ \\
 7 & $\left( \rep{1}; \rep{1},\rep{1}; \crep{5}\right)_{            0}$ & $s_{i}^{0}$ &
 6 & $\left( \rep{1}; \rep{1},\rep{1};  \rep{5}\right)_{            0}$ & $s_{i}^{0}$ \\
 2 & $\left( \rep{1}; \rep{1},\rep{1};\crep{10}\right)_{            0}$ & $s_{i}^{0}$ &
 1 & $\left( \rep{1}; \rep{1},\rep{1}; \rep{10}\right)_{            0}$ & $s_{i}^{0}$ \\
\hline
\end{tabular}
\caption{Massless matter spectrum of the second $\Z{6}$-II model with $\Delta(54)$ flavor symmetry. 
Many SM matter fields build three-dimensional representations of $\Delta(54)$: For instance, all 6 
up-quarks $\bar{u}_i$ combine to two three-dimensional representations of $\Delta(54)$. 
Furthermore, several SM singlets $s_{i}^{0}$ are three-dimensional representations of $\Delta(54)$. 
Finally, the SM singlets $f_{i}$ and $\bar{f}_{i}$ are flavons of $\SU{3}_\mathrm{flavor}$ 
(interestingly, some of these flavons are simultaneously triplets of $\Delta(54)$).}
\label{tab:Z6IIDelta54Model438}
\end{table}

\providecommand{\bysame}{\leavevmode\hbox to3em{\hrulefill}\thinspace}
\frenchspacing
\newcommand{\origttfamily}{}
\let\origttfamily=\ttfamily
\renewcommand{\ttfamily}{\origttfamily \hyphenchar\font=`\-}

\end{document}